\documentclass[aps,prb,twocolumn,showpacs,superscriptaddress]{revtex4}

\usepackage{graphicx}
\usepackage{dcolumn}
\usepackage{amsmath}
\usepackage{amssymb}
\usepackage{rotating}
\usepackage{subfigure,amsmath,verbatim,moreverb}
\def\rv{{\bf r}}
\newcommand{\re}{{\bf r}  }
\newcommand{\rp}{{\bf r'}  }
\newcommand{\kv}{{\bf k}  }
\newcommand{\kf}{k_\mathrm{F}  }
\begin{document}
\title{Nonempirical Density Functionals Investigated for Jellium:
Spin-Polarized Surfaces, Spherical Clusters, and Bulk Linear Response}
\author{Jianmin Tao}
\affiliation{Theoretical Division and CNLS,
Los Alamos National Laboratory, Los Alamos, New Mexico 87545}
\author{John P. Perdew}
\affiliation{Department of Physics and Quantum Theory Group, Tulane
University, New Orleans, Louisiana 70118}
\author{Luis Miguel Almeida}
\affiliation{Department of Physics, University of Aveiro, 
3810 Aveiro, Portugal}
\author{Carlos Fiolhais}
\affiliation{Department of Physics and Center for Computational
Physics, University of Coimbra, 3004-516 Coimbra, Portugal}
\author{Stephan K\"ummel}
\affiliation{Physics Institute, University of Bayreuth, D-95440 Bayreuth, Germany}
\date{\today}
\begin{abstract}
Jellium, a simple model of metals, is a standard
testing ground for density functionals, both for bulk and for 
surface properties. 
Earlier tests show that the Tao-Perdew-Staroverov-Scuseria (TPSS)
nonempirical meta-generalized gradient approximation (meta-GGA) for the
exchange-correlation energy yields more accurate 
surface energies than the local spin density (LSD) approximation
for spin-unpolarized jellium. 
In this study, work functions and surface energies of a jellium metal 
in the presence of ``internal'' and external magnetic fields are 
calculated with LSD, Perdew-Burke-Ernzerhof (PBE) GGA, and 
TPSS meta-GGA and its predecessor, the nearly nonempirical
Perdew-Kurth-Zupan-Blaha (PKZB) meta-GGA, using 
self-consistent LSD orbitals and densities. 
The results show that:
(i) For normal bulk densities, the surface correlation 
energy is the same in TPSS as in PBE, as it should be since TPSS strives to
represent a self-correlation correction to PBE;
(ii) Normal surface density profiles can be
scaled uniformly to the low-density or strong-interaction limit, and 
TPSS provides an estimate for that limit that is consistent with (but
probably more accurate than) other estimates;
(iii) For both normal and low densities,
TPSS provides the same description of surface magnetism as PBE,
suggesting that these approximations may be generally equivalent for 
magnetism. The energies of jellium spheres with up to 106 electrons  
are calculated using density functionals and 
compared to those obtained with Diffusion Quantum Monte Carlo data, 
including our estimate for the fixed-node correction.
Typically, while PBE energies are too low for spheres with more than 
about two electrons, LSD and TPSS are accurate there.  
We confirm that curvature energies are lower in PBE and TPSS than in LSD.
Finally we calculate the linear response of bulk jellium
using these density functionals,
and find that not only LSD but also PBE GGA and TPSS meta-GGA yield a 
linear-response 
in good agreement with that of the Quantum Monte Carlo method,
for wavevectors of the perturbing external potential 
up to twice the Fermi wavevector.
\end{abstract}
\maketitle
\section{Introduction}
Jellium is a simple model of metals. The surface 
properties of jellium can emulate those of real surfaces.
In this popular model the lattice of ions is replaced by a uniform
positive charge density.  Lang and Kohn~\cite{LK70} made 
the first self-consistent calculation of the surface 
properties of this model using
the local spin density (LSD) approximation~\cite{KS} 
for the exchange-correlation (xc) energy 
within the Kohn-Sham density functional formalism~\cite{KS,PK,WK}.
The results, after a correction for lattice effects, 
agreed surprisingly well with  
experiments. This unexpected success initiated the application
of the density functional theory to surfaces.  
A restricted variational calculation by Mahan~\cite{Mahan} gave density
profiles and surface energies very similar to those of Lang and Kohn.

Perdew and Monnier~\cite{PM76,P77,MP78} performed 
a series of self-consistent calculations
of the surface properties of real and jellium metals within LSD.  
Kautz and Schwartz~\cite{KS76} extended the work of Lang and Kohn to the 
spin-polarized case and calculated self-consistently  
several surface properties of jellium polarized 
by a magnetic field uniform inside and zero outside the edge of the 
positive background. 

Over the past few decades, the correct surface energy of jellium has been 
controversial. Recent work~\cite{PP,SSTP2,CPT,PCP,W07,ybs06,CPDGP} resolves  
most of the controversy, placing this energy
close to but probably a little higher than that of LSD
(which benefits from a strong error cancellation between exchange 
and correlation).
For a review of this and earlier work, see Ref.~\onlinecite{CPDGP}.

\par
In density functional theory~\cite{KS,PK,WK}, 
everything is treated exactly except 
the exchange-correlation (xc) energy, which has to be approximated 
as a functional of the electron density. 
Development of better density functional approximations 
has been the subject of continuing theoretical efforts. 
LSD is the simplest approximation and has been 
successful in condensed matter physics. However, it tends to overbind
molecules. An efficient way to solve this problem is to introduce 
the density gradient $\nabla n$ as an additional local ingredient to construct 
a gradient-corrected density functional, the 
generalized gradient approximation (GGA)~\cite{PW86,PBE}.    
With the advent of GGA, density functional theory has become
a popular method in quantum chemistry as well. Although GGAs
have achieved significant improvement over LSD for most properties and
for diverse systems~\cite{SSTP1}, they usually underestimate the surface 
exchange-correlation energy of a spin-unpolarized jellium, for which LSD
is much more accurate~\cite{KPB,SSTP2}. 
\par
Further systematic improvement over GGAs may be made by imposing
additional exact conditions without losing those 
already satisfied by GGA. This can be done by employing the 
kinetic energy densities $\tau_\uparrow(\rv)$ and
$\tau_\downarrow(\rv)$, and/or the Laplacians of the densities
$\nabla^2 n_\uparrow(\rv)$ and $\nabla^2 n_\downarrow(\rv)$, as further 
additional ingredients. Here the kinetic
energy density of electrons with spin $\sigma$ ($\sigma = \uparrow,
\downarrow$) is defined as
\begin{eqnarray}
\tau_{\sigma}(\rv) = \frac{1}{2}\sum_{i}^{\mathrm{occup}}
|\nabla\psi_{i}^{ \sigma}(\rv)|^2,
\label{eq_tau}
\end{eqnarray}
where $\psi_{i}^{ \sigma}(\rv)$ are the occupied Kohn-Sham orbitals, 
which are implicit functionals of the density $n_{\sigma}(\rv)$. 
(Atomic units $\hbar = m = e^2 = 1$ are used throughout 
unless otherwise explicitly stated.)
This family of density functional approximations is called 
meta-GGA~\cite{Perdew85,PKZB,TPSS}. 
\par
While the exact form of the universal functional remains unknown,
many exact constraints on this exact functional have been uncovered.
Thus the more exact constraints a density functional satisfies, 
the closer it is to the exact universal functional. 
Having this in mind, Perdew, Kurth,
Zupan, and Blaha (PKZB)~\cite{PKZB} constructed 
a meta-GGA from first-principles. 
This nearly non-empirical functional satisfies 
two important constraints which can not be satisfied at
the GGA level: It nearly recovers the known fourth-order 
gradient expansion~\cite{SB96} of
the exchange energy in the slowly-varying limit and 
it is free from self-correlation errors. 
\par 
PKZB meta-GGA has impressively 
corrected~\cite{KPB,PKZB,SSTP2} the too-low surface
exchange-correlation energy of GGA functionals, and has successfully 
improved upon LSD and GGAs in thermochemistry 
for molecular atomization energies~\cite{AES}. The defect of PKZB 
is that it contains an empirical parameter in its exchange term, 
which was fitted to molecular atomization energies. Consequently,
PKZB produces too-long bond lengths and some inaccurate properties 
of hydrogen-bonded complexes~\cite{AES,RS,SSTP1}. These failures
 may be attributed to the unbalanced description of PKZB exchange and
correlation for slowly-varying densities and one- or two-electron 
densities, which are the paradigms in condensed matter physics
and in quantum chemistry, respectively.
\par
Starting with the PBE GGA, Tao, Perdew, Staroverov, 
and Scuseria (TPSS)~\cite{TPSS} have 
constructed a nonempirical meta-GGA. While
the formula for TPSS looks a little more complicated 
than the PKZB one, the guiding theory is simple.   
A sound meta-GGA should be able to describe well the paradigm densities  
of condensed matter physics and quantum chemistry. 
By imposing correct constraints, a meta-GGA can be made accurate 
for diverse systems of interest.  The TPSS construction builds many 
additional correct constraints~\cite{SSTP2,PTSS} into a meta-GGA, 
while retaining those that the GGA has already respected.
As a result, this meta-GGA is uniformly
accurate for various properties of diverse systems~\cite{tprscs},
suggesting the correctness of the TPSS philosophy.
\par
For high-density (exchange dominated) systems such as atoms, 
the TPSS is remarkably accurate~\cite{PTSS,SSPT}. 
In the low-density or strong-interaction limit, TPSS recovers 
the PKZB correlation, which is accurate for spin-unpolarized 
densities. 
The relative spin polarization is defined as
\begin{equation}
\zeta=\frac{n_{\uparrow} - n_{\downarrow} }{n},
\label{def_zeta}
\end{equation}
where $ n = n_\uparrow + n_\downarrow$.
Since TPSS correlation strives to
represent a self-correlation correction to PBE, it
should not change PBE correlation for a system with delocalized electrons,
whether spin-polarized or not.
This requirement is satisfied by TPSS through design
for a spin-unpolarized jellium~\cite{SSTP2}. However, we never impose 
this requirement for a spin-polarized density.
Instead we first scale to the low-density limit~\cite{SP,ZTPS,SPL,SPK} 
and there require~\cite{PTSS} the exchange-correlation energy 
to be correctly independent of spin 
for a model uniformly spin-polarized one-electron Gaussian density 
with constant relative spin polarization in the range 
$0 \le |\zeta| \lesssim 0.7$, like LSD and PBE and unlike PKZB.
\par
In this work, we investigate the spin dependence of TPSS correlation and 
find that it is nearly the same as 
that of the PBE for spin-polarized jellium
generated by magnetic fields, implying that TPSS does not
alter the PBE correlation energy for a spin-polarized 
system with delocalized electrons. 
Since TPSS successfully improves on LSD for spin-unpolarized
jellium and has the proper spin dependence, we estimate the 
surface exchange-correlation energy and work function with the TPSS
meta-GGA functional for a spin-polarized jellium in magnetic fields.
An application to the infinite barrier 
model (IBM)~\cite{March1} of metal surfaces is a related 
test but for rapidly-varying densities. The other tests considered 
here are jellium spheres (which sample the surface and curvature energy) 
and the linear response of bulk jellium.
\par 
\section{Density functional approximations}
Density functionals may be ordered by the ``Jacob's 
ladder"~\cite{Perdew2001,prtssc}
of approximations, according to the type of their local
ingredients, whether constructed nonempirically or empirically.
Here we only focus on the all-purpose nonempirical functionals, LSD, PBE and
TPSS, and  the nearly nonempirical PKZB, but not the ones recently developed
for solids and solid surfaces such as AM05~\cite{ann05}, 
Wu-Cohen~\cite{WC06}, and PBEsol~\cite{pbesol}.
The first three rungs of the ladder to be considered here are LSD, PBE, and TPSS. 
(Note that the GGA described in Ref.~\onlinecite{WC06} is constructed 
in part from TPSS  by the 
approximation ${\tilde q}_b \approx 2p/3$ for slowly-varying densities, 
which is a misinterpretation of a statement in Ref.~\onlinecite{TPSS}; 
it is only for $\alpha = 1$, as in the uniform gas, 
and not for $\alpha \approx 1$, 
that ${\tilde q}_b \approx 2p/3$.)
\par
Because the exchange component of a density functional satisfies the
spin scaling relation~\cite{OP}
\begin{eqnarray}
E_{\rm x}[n_\uparrow,n_\downarrow] =
E_{\rm x}[2n_\uparrow]/2 + E_{\rm x}[2n_\downarrow]/2,
\label{eq_scaling}
\end{eqnarray}
where $E_{\rm x}[n] \equiv E_{\rm x}[n/2,n/2]$
and $n(\rv) = n_\uparrow(\rv) + n_\downarrow(\rv)$, we only need 
an exchange functional $E_{\rm x}[n]$ of a spin-unpolarized system. 
An exchange functional also satisfies the uniform coordinate scaling 
requirement~\cite{LP85}
\begin{equation}
E_{\rm x}[n_\gamma] = \gamma E_{\rm x}[n],
\label{eq_exscale}
\end{equation}
where $n_\gamma = \gamma^3n(\gamma\rv)$ is the scaled 
density of $n(\rv)$. These two constraints are the basic requirements
of an exchange functional.
\par
For a spin-unpolarized (closed shell) system, the exchange functionals
of the first three rungs may be
expressed in the form
\begin{eqnarray}\label{enhance}
E_{\rm x}[n] = \int d^3r\ n \epsilon_{\rm x}^{\rm {unif}}(n)
F_{\rm x},
\label{eq_fx}
\end{eqnarray}
where $\epsilon_{\rm x}^{\mathrm{unif}}(n) = -\frac{3}{4\pi}(3\pi^2n)^{1/3}$ 
is the exchange energy per particle of a uniform electron gas 
and $F_{\rm x}$ is the exchange enhancement factor showing
the nonlocality~\cite{PTSS}  
\begin{eqnarray}
F_{\rm x} = 1 + \kappa - \kappa/(1 + x/\kappa),
\label{eq_enhan}
\end{eqnarray}
with $\kappa = 0.804$ and $x \ge 0$. The order of the ladder rungs 
depends upon
the choice of $x$ 
in Eq.~(\ref{eq_enhan}). We have $x = 0$ for LSD, and 
$x=\mu p$ for PBE, where $\mu = 0.21951$, and 
\begin{eqnarray}
p = |\nabla n|^2/[4(3\pi^2)^{2/3}n^{8/3}] = s^2,
\label{eq_ps}
\end{eqnarray}
is the square of the reduced density gradient $s$. For the TPSS and PKZB
meta-GGAs, $x$ is a function of the two variables $p$ and $z$, where
\begin{eqnarray}
z = \tau^{W}/\tau \le 1
\label{eq_zt}
\end{eqnarray}
with $\tau = \sum_\sigma \tau_{\sigma}$ and with
$\tau^{W} = \frac{1}{8}|\nabla n|^2/n$ being the von Weizs\"acker 
kinetic energy density. In the uniform-gas limit, 
all the density functionals above the first rung correctly reduce to LSD.
This uniform-gas limit is the most important requirement~\cite{PTK} for bulk solids 
and surfaces. 
\par
Since the correlation component of a density functional does not 
have such a simple spin scaling relation as the exchange component, 
we have to build the spin dependence into the correlation part. 
The LSD correlation energy has the form
\begin{eqnarray}
E_{\rm c}^{\rm LSD}[n_\uparrow,n_\downarrow] =
\int d^3r \ n \epsilon_{\rm c}^{\mathrm{unif}}(n_\uparrow,n_\downarrow),
\label{eq_eclsd}
\end{eqnarray}
where 
$\epsilon_{\rm c}^{\mathrm{unif}}$ 
is the correlation energy per electron~\cite{PW92} for
a uniform electron gas. The PBE correlation~\cite{PBE,PBY96} 
is based on the LSD correlation 
\begin{eqnarray}
E_{\rm c}^{\rm PBE}[n_\uparrow,n_\downarrow] =
\int d^3r \ n \epsilon_{\rm c}^{\mathrm{PBE}}(n_\uparrow,n_\downarrow,
\nabla n_\uparrow,\nabla n_\downarrow),
\label{eq_ecpbe}
\end{eqnarray}
where $\epsilon_{\rm c}^{\mathrm{PBE}}$ correctly reduces to 
$\epsilon_{\rm c}^{\mathrm{unif}}$
in the uniform-gas limit.
The TPSS correlation~\cite{TPSS,PTSS} is constructed from the PBE correlation, 
\begin{eqnarray}
E_{\rm c}^{\rm TPSS}[n_\uparrow,n_\downarrow] &=&
\int d^3r \ n \epsilon_{\rm c}^{\mathrm{TPSS}}(n_\uparrow,n_\downarrow,
\nabla n_\uparrow,\nabla n_\downarrow,\tau_\uparrow,\tau_\downarrow),
\nonumber \\
& &
\label{eq_ectpss}
\end{eqnarray}
where 
\begin{eqnarray}
\epsilon_{\rm c}^{\mathrm{TPSS}} = \epsilon_{\rm c}^{\mathrm{revPKZB}}
[1 + d \epsilon_{\rm c}^{\mathrm{revPKZB}}(\tau^W/\tau)^3].
\label{eq_ecdens}
\end{eqnarray}
The quantity $\epsilon_{\rm c}^{\mathrm{revPKZB}}$ of Eq.~(\ref{eq_ecdens})
is the revised PKZB correlation
\begin{eqnarray}
\epsilon_{\rm c}^{\mathrm{revPKZB}} & = &
\epsilon_{\rm c}^{\mathrm{PBE}}(n_\uparrow,
n_\downarrow,\nabla n_\uparrow,\nabla n_\downarrow)
 [1 + C(\zeta,\xi) (\tau^{W}/\tau)^2]
\nonumber \\
& & \nonumber \\
& &
- [1 + C(\zeta,\xi)](\tau^{W}/\tau)^2
\sum_{\sigma}\frac{n_\sigma}{n} \tilde\epsilon_{\rm c}^{\sigma},
\label{eq_revpkzb}
\end{eqnarray}
with $\tilde\epsilon_{\rm c}^{\sigma}$ being 
\begin{eqnarray}
\tilde\epsilon_{\rm c}^{\sigma} & = & {\rm max}[
\epsilon_{\rm c}^{\mathrm{PBE}}(n_{\sigma},0,\nabla n_{\sigma},0),
\nonumber \\
& &
\epsilon_{\rm c}^{\mathrm{PBE}}(n_\uparrow,n_\downarrow,
\nabla n_\uparrow,\nabla n_\downarrow)],
\label{eq_maxec}
\end{eqnarray}
where $\zeta$ is the relative spin polarization defined as
$\zeta(\rv) = [n_\uparrow(\rv) -  n_\downarrow(\rv))/n(\rv)]$
and $\xi = |\nabla \zeta|/[2(3\pi^2n)^{1/3}]$~\cite{TPSS,PTSS,WP91}. Here
$C(\zeta,\xi)$ at $\zeta = 0$ and $\xi = 0$
in Eq.~(\ref{eq_revpkzb}) is chosen so that, in the
low-density or strong-interaction limit, TPSS correlation recovers
PKZB correlation, which is accurate~\cite{SPK,PTSS} for spin-unpolarized densities. 
The parameter $d$ in Eq.~(\ref{eq_ecdens}) is 
chosen such that the self-interaction correction should
not alter the surface PBE correlation energy for spin-unpolarized
jellium with delocalized electrons.
The natural construction of the spin-dependent $C(\zeta,\xi)$ would be
to make the TPSS correlation remain the same as the PBE correlation 
for spin-polarized jellium in the achievable  
(and thus energetically important) range of the uniform
bulk relative spin polarization $0 \le \zeta \lesssim 0.7$. However,
$C(\zeta,\xi)$ is designed instead to make TPSS independent
of spin in the low-density limit
for the one-electron Gaussian density and other densities with uniform
$\zeta$ in the range of $0 \le \zeta \lesssim 0.7$
without changing other properties. 
We will show that these two different procedures are essentially equivalent
in the construction of the spin-dependent parameter $C(\zeta,\xi)$.

\section{Spin-polarized jellium and Kohn-Sham approach} 
\begin{table*}
\caption{The strength $\mu_B B_0$ of two model external
magnetic fields of Eqs.~(\ref{eq_mag1}) and~(\ref{eq_mag2}), work
function $W$, and surface correlation and exchange-correlation energies
of the planar jellium surface in 
LSD, PBE, and PKZB and TPSS, as functions of
uniform bulk relative spin polarization $ \zeta$ at normal bulk densities
$ r_s = 2, 4, 6$. $\mu_B B_0$ and $W$ are in eV; surface energies are
in erg/cm$^2$. LSD orbitals and densities are used. 
(1 hartree = 27.21 eV; 1 hartree/bohr$^2$ = $1.557\times 10^6$  erg/cm$^2$)}
\begin{ruledtabular}
\begin{tabular}{llllccccccccc}
 &  & &
& \multicolumn{4}{c}{$\sigma_{\rm c}$} & & \multicolumn{4}{c}{$\sigma_{\rm xc}$}
\\ \cline{5-8} \cline{10-13}
$ r_s$ & $ \zeta$ & $\mu_B B_0$& $W$
& \multicolumn{1}{c}{ LSD} & \multicolumn{1}{c}{PBE} & \multicolumn{1}{c}{PKZB} &
\multicolumn{1}{c}{TPSS} &
& \multicolumn{1}{c}{ LSD} & \multicolumn{1}{c}{PBE} & \multicolumn{1}{c}{PKZB} &
\multicolumn{1}{c}{TPSS}  \\ \hline
 & & & & & & & & & & & \\
 & & & & & & $B(\rv) = B_0\theta(-x)$& & & & & \\
 & & & & & & & & & & & \\
\hline
  &0.0 &0.0 & 3.80 & 317&827& 824& 827 & & 3354 &3264  &3401 & 3380 \\
  &0.2 &1.29 &3.78 &316 &823  &822  &822 & &3350  &3264 &3404 & 3380\\
2 &0.4&2.59 &3.75  &312 &811  & 818&808 & &3337 &3262 &3410  & 3377\\
  &0.6&3.95 &3.68 &301  &785  & 810 &780 & &3315  &3246 &3414 & 3360\\
  &0.8&5.44 &3.54 &275 & 731& 794 & 724 & & 3268 & 3181&3390 &3294 \\
\hline
  &0.0 &0.0 & 2.91&39&124&124&124& &262&253&266&266 \\
  &0.2 &0.26&2.91&40&123&124&123& &261&252&267&265 \\
4 &0.4&0.53&2.90&41&119&123&118& &258&251&270&264 \\
  &0.6&0.81&2.88&44&112&123&110& &252&249&274&260 \\
  &0.8&1.12&2.87&47&100&124&98& &243&241&280& 254 \\
\hline
  &0.0 &0.0 & 2.34&10&40&40&40& &54&52&55&55 \\
  &0.2 &0.10 &2.32&11&39&40&39& &53&51&55&55 \\
6 &0.4&0.20 &2.31&12&38&40&38& &52&51&56&54 \\
  &0.6&0.30 &2.30&14&35&41&35& &50&49&58&52 \\
  &0.8&0.43 & 2.29&17&32&43&32& &47&47&61&51 \\
\hline
 & & & & & & & & & & & \\
 & & & & & & $B(\rv) = B_0$& & & & & \\
 & & & & & & & & & & & \\
\hline
  &0.2 &1.29 &4.21&395&783&794&783& &3363&3469&3620&3586 \\
2 &0.4 &2.59 &5.20&578&706&758&705& &3384&3883&4074&4001 \\
  &0.6 &3.95 &6.46&782&627&733&636& &3466&4324&4571&4454 \\
  &0.8 &5.44 &7.93&907&565&715&591& &3734&4705&4999&4853 \\
\hline
  &0.2 &0.26 &2.95&46&120&121&119& &261&260&276&273 \\
4 &0.4 &0.53 &3.06&64&110&119&108& &258&277&300&289 \\
  &0.6 &0.81 &3.26&86&98&118&97  & &260&299&335&312 \\
  &0.8 &1.12 &3.53&106&87&121&91 & &272&322&372&340 \\
\hline
  &0.2 &0.10 &2.28&12&39&39&39& &53&53&57&56 \\
6 &0.4 &0.20 &2.31&17&36&39&35& &52&55&61&58 \\
  &0.6 &0.30 &2.37&23&33&40&32& &51&57&69&61 \\
  &0.8 &0.43 &2.46&30&29&42&30& &52&60&78&65 \\
\end{tabular}
\end{ruledtabular}
\label{tab_exc}
\end{table*}
In a semi-infinite jellium metal filling the half space
$x < 0$, the uniform positive background density corresponding to the ion lattice 
may be written as~\cite{LK70} 
\begin{eqnarray}
n_{+}(\rv) = \bar n \, \theta(-x),
\end{eqnarray}
where $\theta(-x)$ is a step function and $\bar n$ is the bulk electron
density. In the absence of an external electric field,
the electron density $n(\rv)$ is related to the 
background density via the charge neutralization condition~\cite{P77,KS76}
\begin{eqnarray}
\int d^3r [n(\rv) - n_{+}(\rv)] = 0.
\label{eq_number}
\end{eqnarray}
\par
In the presence of an external magnetic field ${\bf B(\rv)}$,
the self-consistent single-particle Kohn-Sham equation may be written 
as~\cite{KS76}
\begin{eqnarray}
\biggl\{-\frac{1}{2}\nabla^2 + v_{eff}^{\sigma}(\rv)
\biggr\} \psi_{i}^{\sigma}(\rv) =
\epsilon_{i}^{\sigma} \psi_{i}^{\sigma}(\rv),
\label{eq_ks}
\end{eqnarray}
where $v_{eff}^{\sigma}$ is the local effective potential given by
\begin{eqnarray}\label{kspotential}
v_{eff}^{\sigma}(\rv) & = & v_{\rm ext}(\rv) +
\int d^3 r' \ \frac{n(\rv')}{|\rv' - \rv|}
-  \boldsymbol {\sigma}\cdot \mu_B {\bf B}(\rv) \nonumber \\
& & 
+ \ v_{\rm xc}^{\sigma}(\rv).
\label{eq_veff}
\end{eqnarray}
Here $\boldsymbol {\sigma}\cdot{\bf B} = \sigma B$ with
$\sigma = \pm 1$, $\mu_B = e\hbar/2m$ is the Bohr magneton, 
$v_{\rm xc}^{\sigma}(\rv) =
\delta E_{\rm xc}/\delta n_{\sigma}(\rv)$ is the
exchange-correlation potential, and
$v_{\rm ext}(\rv)$ is the external scalar potential 
\begin{eqnarray}
v_{\rm ext}(\rv) = -\int d^3 r' \ \frac{n_{+}(\rv')}{|\rv' - \rv|}.
\end{eqnarray}
$\sigma = +1$ corresponds to electrons with spin 
$\sigma$ parallel to ${\bf B}$ and $-1$ to electrons with spin
$\sigma$ antiparallel to ${\bf B}$. The electron density 
$n_{\sigma}$ may be evaluated from the occupied Kohn-Sham 
orbitals via
\begin{eqnarray}
n_{\sigma}(\rv) = \sum_{i}|\psi_{i}^{ \sigma}|^2 \theta(
\epsilon_{\rm F}^{\sigma} - \epsilon_i^{\sigma}),
\label{eq_den}
\end{eqnarray}
where  
$\epsilon_{\rm F}^{\sigma} = {k_{\rm F}^{\sigma}}^2/2$ with 
$k_{\rm F}^{\sigma} = (6\pi^2 n_{\sigma})^{1/3}$ is the Fermi energy
per electron of spin $\sigma$. Since $v_{eff}^{\sigma}$
of Eq.~(\ref{eq_ks}) depends upon the density $n_{\sigma}$ of 
Eq.~(\ref{eq_den}), Eqs.~(\ref{eq_ks})--~(\ref{eq_den}) must be solved
self-consistently.

Suppose the external magnetic field is uniform for $x < 0$ (within the bulk
metal). Then the bulk density is spin-polarized uniformly with 
position-independent relative spin polarization.
At the Fermi levels, the spin-up and spin-down electrons have the same
chemical potential
\begin{eqnarray}
\mu_{\uparrow} = \mu_{\downarrow},
\label{eq_chem}
\end{eqnarray}
where 
\begin{eqnarray}
\mu_{\sigma}& =& \delta E/\delta n_{\sigma}(\rv) \nonumber \\
&  &
\nonumber \\
& = &
 \frac{\delta T_s}{\delta n_{\sigma}(\rv)} + v_{\rm ext}(\rv) +
u_H(\rv) - \sigma \mu_B B + \ v_{\rm xc}^{\sigma}(\rv). \nonumber \\
&  &
\label{eq_chemsigma}
\end{eqnarray}
Here $T_s[n_{\uparrow},n_{\downarrow}]$ is the Kohn-Sham kinetic energy 
evaluated via Eq.~(\ref{eq_tau}) and $u_H$ is the Hartree potential
which is given by the second term of Eq.~(\ref{eq_veff}). 
The bulk uniform relative spin polarization is determined by the 
strength $\mu_B B$ of the applied external magnetic field via the relation
\begin{eqnarray}
\mu_B B =  (\epsilon_{\rm F}^{\uparrow} - \epsilon_{\rm F}^{\downarrow})/2 
           + [v_{\rm xc}^{\uparrow}( r_s, \zeta) -
              v_{\rm xc}^{\downarrow}( r_s, \zeta)]/2, \
\label{eq_mag}
\end{eqnarray}
where $ r_s = (3/4\pi \bar n)^{1/3}$ is 
the bulk Seitz radius. For convenience,
we put the spin-up and spin-down Fermi levels at zero energy, so that we have
\begin{eqnarray}
v_{eff}^{\sigma}(\rv)  =  - \frac{\delta T_s}{\delta n_{\sigma}(\rv)}. 
\label{eq_flevel}
\end{eqnarray}
Because the positive background is uniform, 
the electron density is also uniform in the
bulk. Kohn-Sham wavefunctions as solutions of one-electron Kohn-Sham 
equation~(\ref{eq_ks}) 
are plane waves and
$v_{eff}^{\sigma}(-\infty) = - \epsilon_{\rm F}^{\sigma}$.
The electrostatic potential is thus given as 
\begin{eqnarray}\label{ues}
v_{\rm es}(-\infty) = -\epsilon_{\rm F}^{\sigma} - 
v_{\rm xc}^{\sigma}(\bar n,  \zeta) + \sigma\mu_B B,
\label{eq_minf}
\end{eqnarray}
where $v_{\rm es} = v_{\rm ext} + u_H$. \par
Near or at the surface, we may write~\cite{LK70,KS76}
\begin{eqnarray}
\psi_{i}^{\sigma}(\rv) = \psi_{k}^{\sigma}(x)e^{i(k_y y + k_z z)},
\end{eqnarray}
where $k$, $k_y$, and $k_z$ are the magnitudes of the
wave vectors along $x$, $y$, and $z$
directions, respectively. The three-dimensional Kohn-Sham 
equation~(\ref{eq_ks}) reduces then to the one-dimensional one~\cite{LK70} 
for $\psi_{k}^{\sigma}(x)$,
\begin{eqnarray}
\biggl\{-\frac{1}{2}\frac{d^2}{dx^2} + v_{eff}^{\sigma}(x)
- v_{eff}^{\sigma}(-\infty)
\biggr\} \psi_{k}^{\sigma}(x) = \epsilon_{\rm k}
\psi_{k}^{\sigma}(x), \nonumber \\
& &
\label{eq_oneks}
\end{eqnarray}
where $\epsilon_{\rm k} = k^2/2$. Solving Eq.~(\ref{eq_oneks})
for $\psi_{k}^{\sigma}(x)$ yields the density $n_{\sigma}$ via
\begin{eqnarray}
n_{\sigma}(x) = 3\bar n_{\sigma} \int_0^1 d \tilde k \ 
(1 - {\tilde k}^2)
|\psi_{k}^{\sigma}(x)|^2, 
\label{eq_rho}
\end{eqnarray}
where $0\le \tilde k = k/k_F^{\sigma} \le 1$. In the vacuum, 
the density decays exponentially~\cite{MP78}, as in an atom~\cite{TAO01}, 
\begin{eqnarray}
\psi_{k}^{\sigma}(x) \rightarrow  e^{-a x}, \hspace{0.5cm}
(x \rightarrow \infty),
\end{eqnarray}
where $a$ is a constant for a given $\bar n$. \par
Within LSD, the exchange-correlation potential~\cite{PW92} may be evaluated as
\begin{eqnarray}
v_{\rm xc}^{\sigma} =
\frac{\partial}{ \partial n_{\sigma}}
[n \epsilon_{\rm xc}(n_{\uparrow},n_{\downarrow})].
\end{eqnarray}
Following Monnier and Perdew~\cite{MP78} for the 
treatment of the electrostatic potential 
and the self-consistency procedure (outlined in 
Appendix A of Ref.~\onlinecite{MP78}), 
we solved the one-dimensional Kohn-Sham 
equation~(\ref{eq_oneks}) self-consistently within LSD.

In the present work, two external magnetic fields 
coupled to the electron spins are 
considered: \par
(1) Uniform inside and zero outside the jellium edge,
\begin{eqnarray}\label{Kautz}
B(x) = B_0\theta(-x); 
\label{eq_mag1}
\end{eqnarray}
\par
(2) Uniform everywhere,
\begin{eqnarray}\label{ew}
B(x) = B_0.
\label{eq_mag2}
\end{eqnarray}
Eq.~({\ref{Kautz}}) was proposed by Kautz and Schwartz~\cite{KS76} to simulate,
within the jellium model, the "internal" magnetic field near the surface of a
ferromagnetic metal, while Eq.~({\ref{ew}}) can be realized experimentally
over a range of $B_0$. 
The magnitude of the external magnetic fields $B_0$ may be found
from Eq.~(\ref{eq_mag}) for a given $ \zeta$.

\begin{table}
\caption{Spin dependences of the surface exchange-correlation energies
(in units of erg/cm$^2$) of the planar jellium surface in
PBE and TPSS with various bulk valence-densities.
LSD orbitals and densities are used.}
\begin{ruledtabular}
\begin{tabular}{lclddddd}
 &\multicolumn{1}{c}{} & &
\multicolumn{4}{c}{$[\sigma_{\rm xc}( r_s, \zeta) -
\sigma_{\rm xc}( r_s,0)]$} \\
 $ r_s$ & $\sigma_{\rm xc}^{\rm TPSS}( r_s,0)$ & &
0.2 &
0.4 & 0.6 &
0.8 \\ \hline
&&&&&&& \\
& & \multicolumn{4}{c}{$B(\rv) = B_0\theta(-x)$} \\
&&&&&&& \\
2 & 3380 & PBE &0 &-2 &-18 & -83\\
  &        & TPSS&0 &-3 &-20 & -86\\
4 & 266& PBE&-1 &-2 &-4 & -12 \\
  &      &TPSS&-1 &-2 &-6 & -12 \\
6 & 55& PBE &-1  &-1 &-3  & -5 \\
  &      &TPSS &0  &-1 &-3  &-4  \\
&&&&&&& \\
& & \multicolumn{4}{c}{$B(\rv) = B_0$} \\
&&&&&&& \\
2 & 3380 &PBE &205 &619 &1060 & 1441 \\
 &         &TPSS&206 & 621&1074  & 1473 \\
4 & 266 & PBE &7  &24  &46  &69  \\
  &       &TPSS &7  &23  &46  & 74 \\
6 & 55 & PBE &1  &3  &5  &8  \\
  &       & TPSS&1  &3  &6  &10  \\
\end{tabular}
\end{ruledtabular}
\label{tab_norm}
\end{table}
\begin{table}
\caption{Spin dependences of the surface exchange-correlation energies
of the planar jellium surface in
PBE and TPSS, when the enhancement factor $F_{\rm xc}$ of 
Eq.~(\ref{xcenhance}) is 
uniformly scaled to the high-density
($r_s \rightarrow 0$) or exchange-only limit
from normal bulk valence-densities.
LSD orbitals and densities are used. (erg/cm$^2$)}
\begin{ruledtabular}
\begin{tabular}{lclddddd}
 &\multicolumn{1}{c}{$\sigma_{\rm x}^{\rm TPSS}( r_s,0)$} & &
\multicolumn{4}{c}{$[\sigma_{\rm x}( r_s, \zeta) -
\sigma_{\rm x}( r_s,0)]$} \\
 $ r_s$ & (erg/cm$^2$) & &
0.2 &
0.4 & 0.6 &
0.8 \\ \hline
&&&&&&& \\
& & \multicolumn{4}{c}{$B(\rv) = B_0\theta(-x)$} \\
&&&&&&& \\
2 & 2553 &PBE &5 &15 &25 &14  \\
 &         &TPSS&5 &16 &26  &16  \\
4 & 141 & PBE &1  &4  &9  &13  \\
  &       &TPSS &2  &5  &9  &15  \\
6 & 15 & PBE &0  &1  &2  &3  \\
  &       & TPSS&1  &2  &3  &4  \\
&&&&&&& \\
& & \multicolumn{4}{c}{$B(\rv) = B_0$} \\
&&&&&&& \\
2 & 2553 &PBE &250 &740 &1261 &1704  \\
 &         &TPSS&250 &742 &1265  &1709  \\
4 & 141 & PBE &12  &39  &73  &107  \\
  &       &TPSS &13  &40  &74  &108  \\
6 & 15 & PBE &2  &7  &13  &19  \\
  &       & TPSS&3  &7  &14  &15  \\
\end{tabular}
\end{ruledtabular}
\label{tab_high}
\end{table}
The work function $W$~\cite{LK70,KS76,MP78} 
is an interesting quantity which can be 
measured experimentally. It is defined as the energy required to 
remove an electron from a bulk solid into the vacuum.
Within the framework of Kohn-Sham density functional theory the work 
function is the
difference of the Kohn-Sham single-particle energies 
of an electron at rest in the vacuum and an electron moving
at the Fermi level in the bulk~\cite{MP78}, {\it i.e.},  
\begin{eqnarray}
W = v_{\rm eff}^{\sigma}(\infty) - 
[ {k_{\rm F}^{\sigma}}^2/2 + v_{\rm eff}^{\sigma}(-\infty)].
\end{eqnarray} 
In the presence of an exernal magnetic field, the work function 
is given by~\cite{KS76}
\begin{eqnarray}
W &=& v_{\rm eff}^{\sigma}(\infty) - 
[ {k_{\rm F}^{\sigma}}^2/2 + v_{\rm eff}^{\sigma}(-\infty)]
+ \sigma \mu_B B \nonumber \\
& = & [v_{\rm es}(\infty) - v_{\rm es}(-\infty)] - 
v_{\rm xc}^{\sigma}(\bar n_{\uparrow},\bar n_{\downarrow}) - 
{k_{\rm F}^{\sigma}}^2/2 \nonumber \\
& &
+ \sigma \mu_B B,
\end{eqnarray}
where the second equality can be obtained by combining 
Eqs.~(\ref{kspotential}) and~(\ref{ues}).

\section{Surface energies of spin-polarized jellium} 
\subsection{Surface energies}
The surface energy of a solid is the energy required to split
the solid per unit area of new surface formed. Here we only
focus on the exchange-correlation component 
$\sigma_{\rm xc}$ of the 
total surface energy. The surface exchange-correlation energy
is 
\begin{eqnarray}
\sigma_{\rm xc} = \int_{-\infty}^{\infty}dx \
n(x)[\epsilon_{\rm xc}(x) - \epsilon_{\rm xc}(-\infty)],
\end{eqnarray}
and may be decomposed as a sum of the exchange and 
correlation contributions $\sigma_{\rm xc} = \sigma_{\rm x} +
\sigma_{\rm c}$. \par
We evaluated the surface 
exchange and correlation energies
of spin-polarized jellium produced by the two model external magnetic fields
of Eqs.~(\ref{eq_mag1}) and~(\ref{eq_mag2}). The results are
displayed in Table~\ref{tab_exc}. The surface exchange and 
correlation energies of spin-unpolarized jellium are also
listed for comparison.
\begin{table}
\caption{Spin dependences of the surface exchange-correlation energies
of the planar jellium surface in 
PBE and TPSS, when the enhancement factor $F_{\rm xc}$ of
Eq.~(\ref{xcenhance}) is
uniformly scaled to the low-density
($r_s \rightarrow \infty$) limit
from normal bulk valence-densities.
LSD orbitals and densities are used. 
The $\zeta$-dependence here arises not so much from
$F_{\rm xc}$ as from the $\zeta$-dependence of the surface density 
profile $n(x)$, which for $B(\rv) = B_0$
spreads out more as $|\zeta|$ increases.
(erg/cm$^2$)}
\begin{ruledtabular}
\begin{tabular}{lclddddr}
 &\multicolumn{1}{c}{$\sigma_{\rm xc, \infty}^{\rm TPSS}( r_s,0)$} & &
\multicolumn{4}{c}{$[\sigma_{\rm xc, \infty}( r_s, \zeta) -
\sigma_{\rm xc, \infty}( r_s,0)]$} \\
 $ r_s$ & (erg/cm$^2$) & &
0.2 &
0.4 & 0.6 &
0.8 \\ \hline
&&&&&&& \\
& & \multicolumn{4}{c}{$B(\rv) = B_0\theta(-x)$} \\
&&&&&&& \\
2 & 6364 & PBE &-17 &-77 &-215 &-504 \\
  &        & TPSS&-19 &-81 &-220 &-489 \\
4 & 503& PBE&1 &3 &2 &-10  \\
  &      &TPSS&1 &1 &-2 &-7  \\
6 & 107& PBE &1  &3 &5  &3  \\
  &      &TPSS &1  &2 &4  &6  \\
&&&&&&& \\
& & \multicolumn{4}{c}{$B(\rv) = B_0$} \\
&&&&&&& \\
2 & 6364 & PBE &477 &1461 &2352 &2714 \\
  &        & TPSS&470 &1441 &2352 &2753 \\
4 & 503& PBE&26 &86 &149 &180  \\
  &      &TPSS&25 &83 &148 &193  \\
6 & 107& PBE &6  &18 &31  &37  \\
  &      &TPSS &6  &17 &31  &42  \\
\end{tabular}
\end{ruledtabular}
\label{tab_low}
\end{table}
In our calculations, we employed LSD orbitals and densities obtained by 
self-consistently solving the Kohn-Sham equation~(\ref{eq_oneks}) 
using the LSD exchange-correlation potential. For a
justification of this approach, see Ref.~\onlinecite{apf}. 
The work functions shown in Table I are LSD values.

Previous studies~\cite{SSTP2} show that, like PKZB,
 TPSS successfully improves upon LSD
in the surface exchange-correlation energy of a   
spin-unpolarized jellium, while PBE gives a correction of the wrong sign
to LSD and thus underestimates this quantity. Figures 1 and 2
clearly show that, while the surface exchange energy of LSD is much more 
overestimated than those of the PBE GGA and TPSS meta-GGA,
the surface exchange-correlation energies of these density functionals
are not very different from each other. 
Fig. 2 shows that the improvement of
TPSS over LSD and PBE may be attributed to
its recovery of the known correct gradient expansions of
the exchange~\cite{SB96,TPSS}-correlation~\cite{MB68,GR76,LP80} energy. 
Pictured in Fig. 2 is the dependence of
the exchange-correlation enhancement factor,
\begin{eqnarray}\label{xcenhance}
F_{\rm xc} = 
\epsilon_{\rm xc}^{\mathrm{TPSS}}(n_\uparrow,n_\downarrow,
\nabla n_\uparrow,\nabla n_\downarrow,\tau_\uparrow,\tau_\downarrow)/
\epsilon_{\rm x}^{\rm unif}(n), 
\end{eqnarray} 
upon position $x$.

\begin{figure}
\includegraphics[width=\columnwidth]{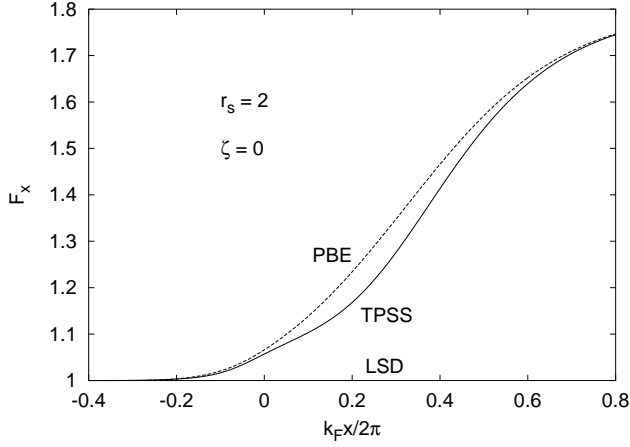}
\caption{Exchange enhancement factor $F_{\rm x}$ of Eq.~({\ref{enhance}})
for LSD, PBE, and TPSS as functions of position
$x$ in units of $2\pi/ k_{\rm F}$ relative to the jellium edge ($x = 0$)
for spin-unpolarized jellium with $ r_s = 2$.}
\end{figure}
\begin{figure}
\includegraphics[width=\columnwidth]{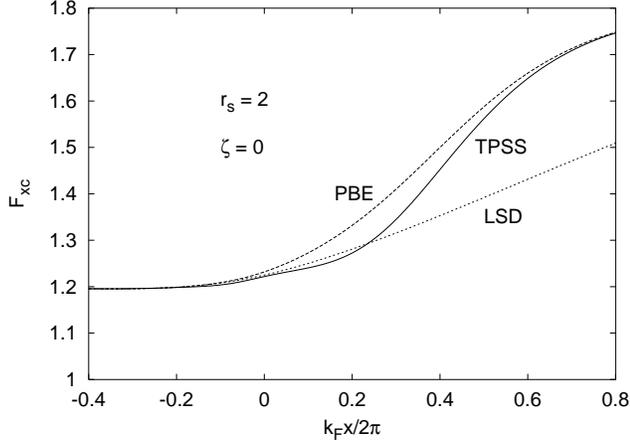}
\caption{Exchange-correlation enhancement factors $F_{\rm xc}$
of Eq.~({\ref{xcenhance}}) for LSD, PBE, and TPSS as functions of
position $x$ in units of $2\pi/ k_{\rm F}$ relative to
the jellium edge ($x = 0$)
for spin-unpolarized jellium with $ r_s = 2$.
The same (LSD) surface density profile $n(x)$ is assumed for all
three functionals. Thus the differences in $F_{\rm xc}(x)$ shown here
determine the differences in surface exchange-correlation energy
$\sigma_{\rm xc}$ seen in the $\zeta = 0$ row of Table I.  The higher is
$F_{\rm xc}$ at a given x, the lower is $\sigma_{\rm xc}$.
But the large-$x$ tail region of low density is much less important
than the region close to the edge of the positive background at $x = 0$.
}
\end{figure}
\begin{figure}
\includegraphics[width=\columnwidth]{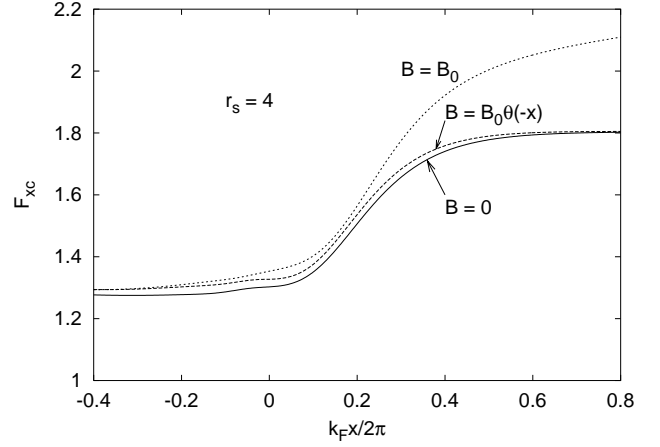}
\caption{Comparison of TPSS exchange-correlation enhancement 
factor $F_{\rm xc}$ of Eq.~({\ref{xcenhance}}) as a function of $x$ 
in units of $2\pi/ k_{\rm F}$ relative to
the jellium edge ($x = 0$), for spin-unpolarized jellium at 
$ r_s = 4$, and for spin-polarized jellium at $ r_s = 4$ and
$ \zeta = 0.4$ produced by two external 
magnetic fields of Eqs.~(\ref{eq_mag1}) and~(\ref{eq_mag2}).}
\end{figure}
\begin{figure}
\includegraphics[width=\columnwidth]{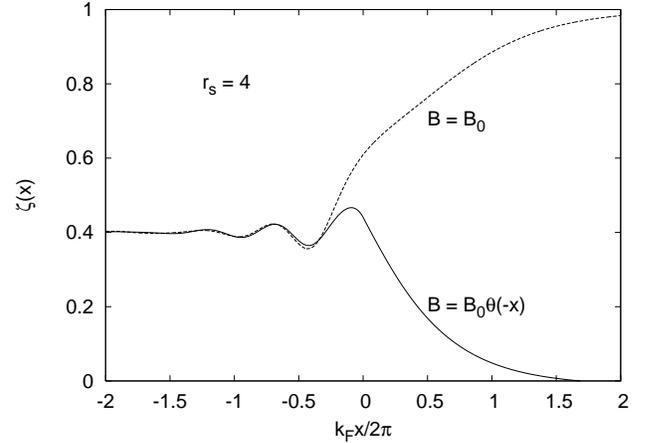}
\caption{Relative spin polarization 
$\zeta = (n_\uparrow - n_\downarrow)/n$ as a function of $x$
in units of $2\pi/ k_{\rm F}$ relative to
the jellium edge ($x = 0$) for spin-polarized jellium at $ r_s = 4$ and
$ \zeta = 0.4$ produced by two external
magnetic fields of Eqs.~(\ref{eq_mag1}) and~(\ref{eq_mag2}).}
\end{figure}

Table I suggests that the work function and surface exchange-correlation
energy of a metal could be slightly reduced by an increase in bulk spin
polarization due to ferromagnetism [$B = B_0\theta(-x)$], but could be
strongly increased by an increase in spin polarization due to an
external uniform magnetic field.
For the work function of the model ferromagnet, our results agree
qualitatively with those of Kautz and Scwartz~\cite{KS76}
(who however used random phase approximation (RPA) input to
their LSD calculation).

The local spin polarization $\zeta(x)$ varies from its bulk value at
$x=-\infty$ to a limiting tail value $\zeta(\infty)$
at $x=+\infty$.  For $B = B_0\theta(-x)$, $\zeta(\infty) = 0$, but for
$B = B_0$, $\zeta(\infty) = 1$. These different limits are reflected
in Fig. 3, a plot of exchange-correlation enhancement factors vs. position $x$,
and are presented directly in Fig. 4.

\subsection{Spin dependences of PBE GGA and TPSS meta-GGA}
To examine the spin dependences of the PBE GGA and TPSS meta-GGA 
for a spin-polarized jellium, we define the surface
exchange-correlation spin-polarization energy as
\begin{eqnarray}
\Delta \sigma_{\rm xc} = \sigma_{\rm xc}( r_s, \zeta)
- \sigma_{\rm xc}( r_s, 0).
\label{eq_change}
\end{eqnarray}
This quantity $\Delta \sigma_{\rm xc}$ can tell us how the
surface exchange-correlation energy of a spin-polarized jellium
changes when $ \zeta$ changes. Tables~\ref{tab_norm},~\ref{tab_high},
and~\ref{tab_low} 
show the comparison of the spin dependences of PBE and TPSS
for normal bulk valence densities and when the enhancement factor 
$F_{\rm xc}$ of Eq.~(\ref{xcenhance}) is
uniformly scaled to 
the high-density ($r_s \rightarrow 0$ or exchange-only) and 
low-density~\cite{SPK,PTSS} ($r_s \rightarrow \infty$ 
or strong-interaction) limits,
respectively. From Tables~\ref{tab_norm},~\ref{tab_high}, 
and~\ref{tab_low} we see that
TPSS has nearly the same dependence as PBE in every case 
we examined here, suggesting that our earlier procedure to construct
the spin-dependence of $C(\zeta,\xi)$ in Eq.~(\ref{eq_revpkzb}) is
right.

The PBE GGA and TPSS meta-GGA describe magnetism very similarly at 
jellium surfaces. Whether this will remain true for real systems remains 
to be determined.  We only know of a study of the ground-state
spins of iron complexes~\cite{SGEL}, which compares PBE with TPSSh~\cite{SSTP1} 
(a hybrid of TPSS with $10\%$ exact exchange), and
another study of iron complexes~\cite{ZBFCC} which seems to show similar 
energy gaps between high- and low-spin states from PBE and TPSS. 
A private communication~\cite{Swart}
from one of the authors of Ref.~\onlinecite{SGEL} shows that PBE and TPSS give 
similar energy differences among three spin states in each of 
seven iron complexes.
\begin{table}
\caption{Surface exchange-correlation energy of jellium in
the infinite barrier model. March's unit
$3 e^2 \bar n/4\pi = (88738/r_s^3)({\rm erg/cm^2})$ is used~\cite{March1}. 
The RPA+ value for $r_s = 0$ is from Ref.~\onlinecite{VS1}.} 
\begin{ruledtabular}
\begin{tabular}{llllccccccccc}
$ r_s$ 
& \multicolumn{1}{c}{ RPA +}
& \multicolumn{1}{c}{ LSD} & \multicolumn{1}{c}{PBE} & \multicolumn{1}{c}{PKZB} &
\multicolumn{1}{c}{TPSS}  \\ \hline
0 &0.0714 &0.1108  & 0.0452& 0.0501&0.0516  \\
2 &0.1289 &0.1222 & 0.1035 & 0.1079 & 0.1098 \\
4 &0.1364 &0.1296 &0.1190 & 0.1235 & 0.1255 \\
6 &0.1393 &0.1350 &0.1286 &0.1335 & 0.1354 \\
$\infty$&...&0.2157 &0.2313 &0.2588 &0.2603 \\
\end{tabular}
\end{ruledtabular}
\label{tab_ibm}
\end{table}
\section{Infinite barrier model of the jellium surface}
The earliest surface model of a metal is the infinite barrier model
proposed by Bardeen~\cite{Bardeen},
in which the surface density profile is that of noninteracting
free electrons in the presence of a hard wall.  
This is an oversimplified surface model of a jellium metal,  
because many properties of this model surface can not be
transferred to real metal surfaces.
However, it may serve as an ideal model system with rapidly varying
densities and can be employed to test density functionals, 
because the exact electron density, the kinetic energy density, 
and the conventional
exchange energy per electron of this model system are analytically 
known~\cite{March1, March2}. Here, the surface exchange-correlation 
energies of the LSD, PBE, PKZB and TPSS are evaluated for normal 
bulk valence densities,
and when the enhancement factor
$F_{\rm xc}$ of Eq.~(\ref{xcenhance}) is
uniformly scaled to the high-density or exchange-only and 
low-density or strong-interaction limits. The results are shown
in Table~\ref{tab_ibm}.

RPA+ is a sophisticated approximation involving the full 
random phase approximation (RPA) plus
a GGA for the short-range correction to RPA. 
The RPA+ values in Table~\ref{tab_ibm}
are evaluated from~\cite{LP00}
\begin{eqnarray}
\sigma_{\rm xc}^{\rm RPA+} = 0.0714(1 + 3.451r_s)/(1 + 1.688r_s),
\label{eq_rpa}
\end{eqnarray}
which is a fit to the RPA+ values of Yan, Perdew, 
and Kurth~\cite{YPK}
at $r_s = 0$, 2.07, 4 and 6. The RPA+ values are exact in the 
exchange-only or $r_s \rightarrow 0$ limit, and we take them to 
be nearly exact for all rs.  
We see from Table~\ref{tab_ibm} that
the order of accuracy of these functionals for normal bulk valence 
densities is
 $\sigma_{\rm xc}^{\rm PBE} <
\sigma_{\rm xc}^{\rm PKZB} \approx \sigma_{\rm xc}^{\rm TPSS} <
\sigma_{\rm xc}^{\rm LSD}$, while, in the high-density or 
exchange-only limit, we have 
$\sigma_{\rm x}^{\rm LSD} < \sigma_{\rm x}^{\rm PBE} <
\sigma_{\rm x}^{\rm PKZB} \approx \sigma_{\rm x}^{\rm TPSS}$.
\begin{figure}
\includegraphics[width=\columnwidth]{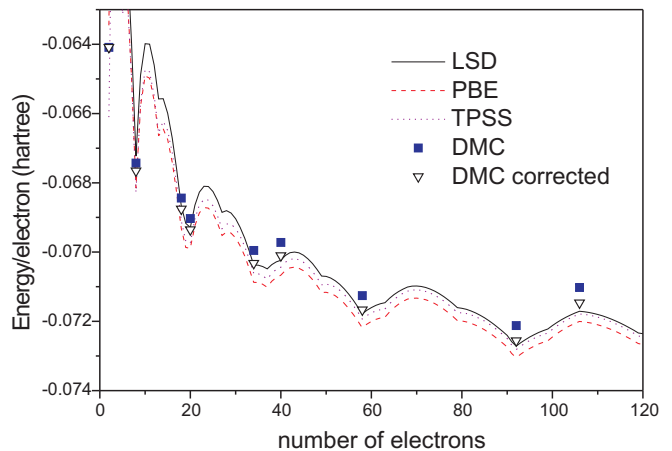}
\caption{Total energies per electron of jellium spheres 
for $r_s=4$. The effect of our fixed-node correction is visible for
magic clusters $N=$ 2, 8, 18, 20, 34, 40, 58, 92, and 106. 
PBE is too low, but TPSS and especially LSD are good for $N>2$.
LSD is best for $N>2$, but worst for $N=2$.  
Although TPSS surface energies are slightly higher than LSD values, 
TPSS curvature energies (Fig. 6 and Table IX) are lower, leading 
to a slightly lower TPSS total energy for these spheres.
All curves tend as $N \rightarrow \infty$ to $-0.0774$ hartree.}
\label{corrdmc}
\end{figure}
\begin{figure}
\includegraphics[width=8.5cm]{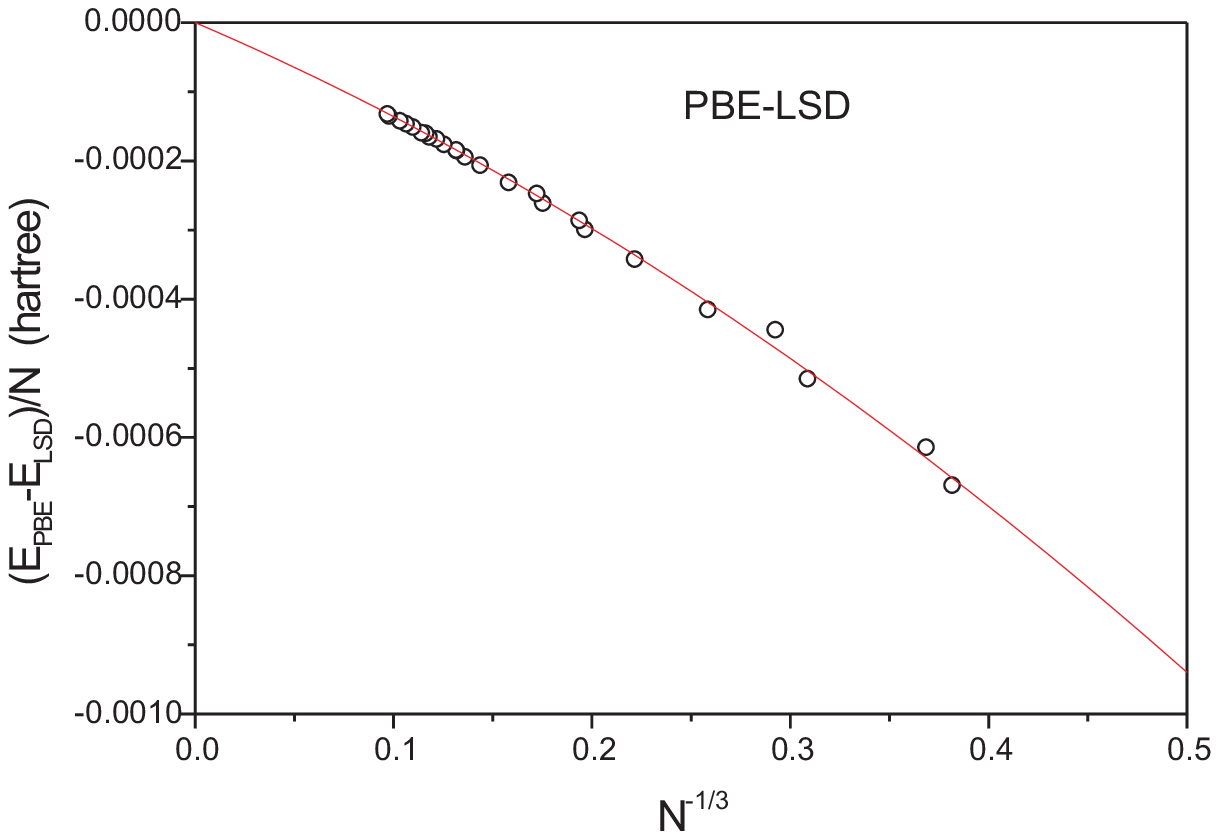}
\includegraphics[width=8.5cm]{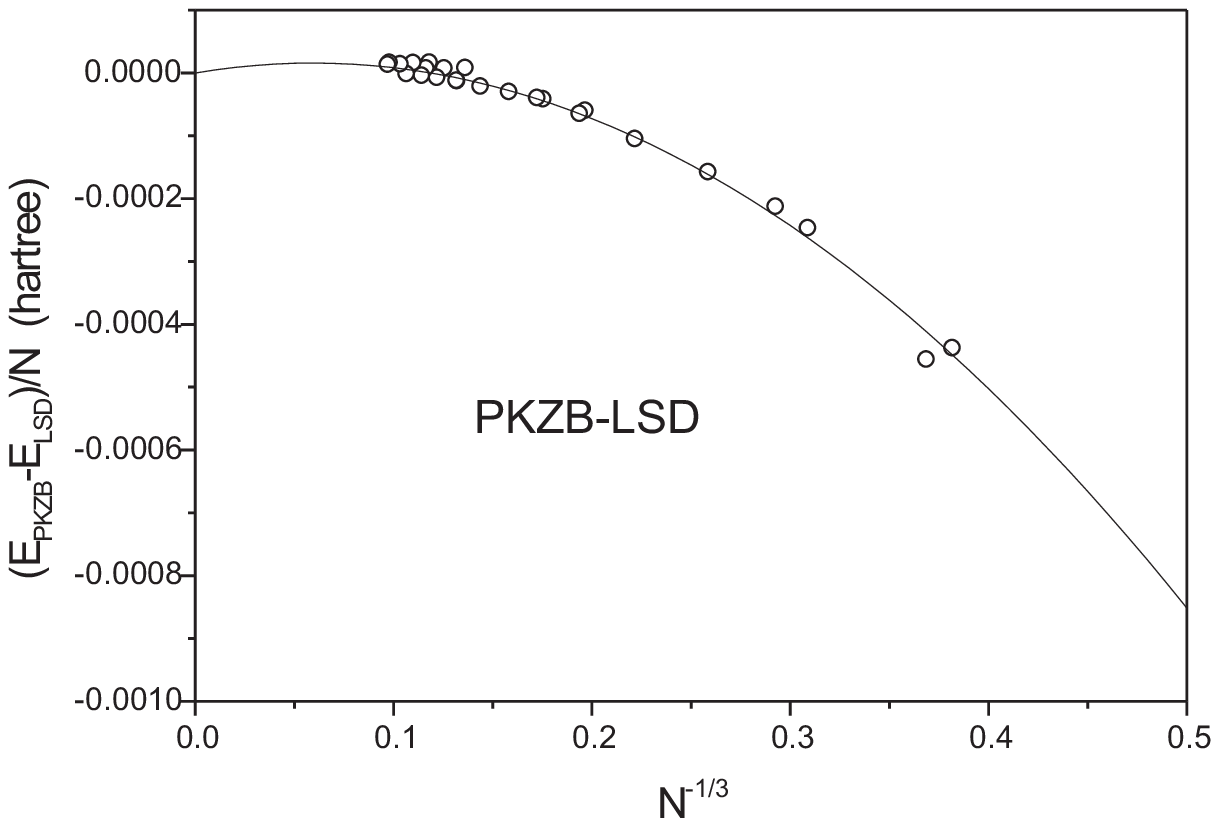}
\includegraphics[width=8.5cm]{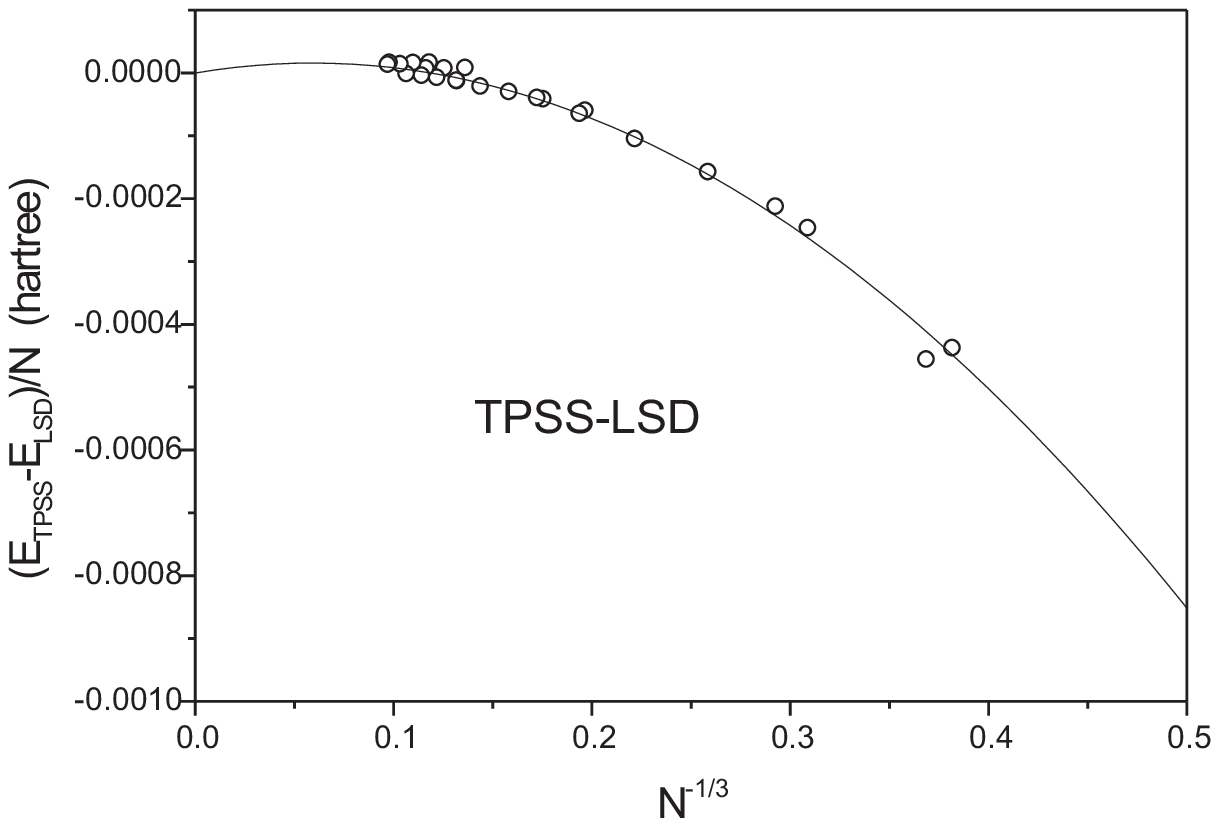}
\caption{Deviation from LSD of PBE, PKZB and TPSS energies for
jellium spheres with $r_s$ = 3.93. The full lines are parabolas
fitted to the LDM via Eq.~(\ref{eqdiffldmn}), as explained in the text.
The open circles are input values for the magic clusters respecting
the ``Aufbau" principle. 
The second derivative of this curve at $N^{-1/3} = 0$ is 
proportional to the difference of curvature energies between 
the given functional and LSD.}
\label{fit400}
\end{figure}
\section{ENERGIES OF JELLIUM SPHERES}

In the spherical jellium model a positive charge background is 
contained inside a sphere of radius  
\[ R=r_s N^{1/3}.\]
The potential due to the positive background charge is
\begin{equation}
v_{+}(r)=\left\{ 
\begin{tabular}{lll}
$-\frac{N}{2R}\left[3-\left(\frac{r}{R}\right)^2\right]$ &  & $(r\leq R)$ \\
&  &  \\ 
$-\frac{N}{r}$ &  & $(r>R)$.
\end{tabular}
\right.
\end{equation}
Calculation of the exchange and correlation energies was done 
in an {\it a posteriori} process using LSD densities as input, as explained 
in Ref. \onlinecite{apf}. By solving the many-electron problem using 
the Kohn-Sham approach we may obtain a series of single particle levels. 
For metallic densities the energy ordering of these states is 
1s, 1p, 1d, 2s, 1f, 2p, 1g, 2d, 1h, 3s, 2f, ~ ... The filling of shells 
yields special stability 
for the so-called magic clusters.

As in  Ref. \onlinecite{apf}, we calculated the energies of 
jellium spheres for some magic numbers ($N=$ 2, 8, 18, 20, 34, 
40, 58, 92 and 106) with various densities $r_s$ = 2.07, 3.25, 
4.00 and 5.62. We chose these magic numbers and densities due to 
the availability of the respective results in 
Diffusion Quantum Monte Carlo 
calculations by Sottile and Ballone \cite{sb}, with which 
we want to compare. These Diffusion Quantum Monte Carlo (DMC) 
results on jellium spheres are supposed to be the most accurate 
available for the systems under study. However, Sottile-Ballone 
values\cite{sb} are affected by a systematic error in the 
correlation energies due to their fixed-node assumption.

For $N\rightarrow\infty$ we get the limit of the uniform electron gas 
for which the fixed-node error may be estimated as the difference 
between the fixed-node calculation done by Ortiz-Ballone \cite{ob94} 
and the released-node calculation of Ceperley-Alder \cite{ca80}. 
This error estimate for the uniform electron gas has been presented 
in Table VII of Ref.~\onlinecite{apf}. 
For a sphere with $N=2$ electrons, 
there is no node in the space factor of the ground-state wavefunction 
so the fixed-node error is absent. Using the 
limits $N=\infty$ and $N=2$ we may interpolate the fixed-node error of 
the intermediate spheres by multiplying the correction to the 
correlation energy in the uniform electron gas by a  
factor suggested by the Liquid Drop Model~\cite{pwe} (Eq.~(\ref{eqldmn})
below): 
\begin{equation}
\Delta\epsilon_{c}(N)=
\Delta\epsilon^{unif}_{c}\left[1-\left(\frac{2}{N}\right)^{1/3}\right].
\label{eccorrected}
\end{equation}
The uniform electron gas correction $\Delta\epsilon^{unif}_{c}$ is 
equal to the $\Delta\epsilon^{OB}_{c}$ of Table VII of Ref. \onlinecite{apf}.

A better comparison of the energetics produced by the various density 
functionals can now be done. The effect of correction is shown in 
Fig.~\ref{corrdmc}  for a single density ($r_s$=4.0 bohr).
The error of a density functional is the 
difference between the exchange-correlation energy and the corresponding 
corrected DMC value. Table~\ref{totdev} shows the errors of the 
different density functionals for the total energy per electron, 
and for the five indicated densities, averaged over the magic 
closed-shell clusters in the range $2\leq N\leq 106$. 
\begin{table}
\caption{Mean absolute deviations from corrected fixed-node DMC values
[Ref.~{\protect\onlinecite{sb}} and Eq.~(\ref{eccorrected})] of the total
energies per electron of jellium spheres in various density
functional approaches. The values are averages over the magic clusters
$N=$  2, 8, 18, 20, 34, 40, 58, 92, and 106. For individual $N$ at $r_s = 4$, 
see Fig. 5.}
\begin{tabular}{c c c c c }
\hline \hline                                   
\vspace{3pt}&\multicolumn{4}{c}{$|(E-E^{DMC})/N|$ (hartree)} \\
~~~$r_s$~~~ & ~~~LSD~~~ & ~~~PBE~~~ & ~~~PKZB~~~ & ~~~TPSS~~~   \\
\hline 
1.00    &       0.0040  &       0.0015  &       0.0015  &       0.0007  \\
2.00    &       0.0018  &       0.0007  &       0.0004  &       0.0004  \\
3.25    &       0.0007  &       0.0005  &       0.0002  &       0.0004  \\
4.00    &       0.0004  &       0.0005  &       0.0004  &       0.0005  \\
5.62    &       0.0005  &       0.0006  &       0.0006  &       0.0007  \\
\hline 
average&        0.0015  &       0.0008  &       0.0006  &       0.0006  \\
\hline \hline 
\end{tabular}
\label{totdev}
\end{table}

\begin{table}
\caption{Average relative deviations of the correlation energy 
of jellium spheres, in various density functional approaches, 
from corrected DMC values [Ref.~{\protect\onlinecite{sb}} and
Eq. (\ref{eccorrected})]. Averages were taken over magic clusters:
$N=$ 2, 8, 18, 20, 34, 40, 58, 92, and 106.}
\begin{tabular}{c c c c c }
\hline \hline                                   
\vspace{3pt} &\multicolumn{4}{c}{$(E_c-E_c^{DMC})/E_c^{DMC}$ }\\
~~~$r_s$~~~ & ~~~LSD~~~ & ~~~PBE~~~ & ~~~PKZB~~~ & ~~~TPSS~~ \\
\hline
1.00    & 40.3\%        &       6.7\%   &       7.4\%   &       5.9\%   \\
2.00    & 34.0\%        &       7.7\%   &       7.8\%   &       6.5\%   \\
3.25    & 29.7\%        &       7.3\%   &       6.9\%   &       5.8\%   \\
4.00    & 27.2\%        &       6.4\%   &       5.9\%   &       4.9\%   \\
5.62    & 26.4\%        &       7.3\%   &       6.4\%   &       5.5\%   \\
\hline
average& 31.5\% &       7.1\%   &       6.9\%   &       5.7\%   \\
\hline\hline
\end{tabular}
\label{corrDFTvsMC}
\end{table}
Table~\ref{corrDFTvsMC}
 displays the relative errors in the correlation energies, again 
averaged over the closed-shell spheres. The improvement of all functionals 
with respect to LSD is clear and the TPSS shows the smallest deviation 
in correlation energy.

Surface and curvature energies are relevant~\cite{pwe} not only to clusters 
and voids, but even to cohesive energies and monovacancy formation energies.
Adopting the same fitting procedure for extracting Liquid Drop 
Model (LDM)~\cite{pwe} parameters 
from jellium spheres as in Ref. \onlinecite{apf}, we use the following equation 
for the energy of a neutral jellium cluster with $N$ valence electrons
\begin{equation}
\frac{E^{LDM}}{N}=
\epsilon^{unif} + 4\pi r_s^2 \sigma N^{-1/3}+2\pi r_s \gamma N^{-2/3},
\label{eqldmn}
\end{equation}
\noindent where $\sigma$ and $\gamma$ describe the surface and curvature 
energies respectively, and $\epsilon^{unif}=(4\pi r_s^3/3)\alpha$ is the 
energy per electron of the uniform electron gas, with $\alpha$ being 
its energy per volume. This model neglects the quantum oscillations in 
the energy due to the shell structure. For the sequence of closed-shell 
clusters, the oscillation is presumably the same in LSD as at any higher 
level of theory, so the difference cancels out.
(More generally, $[E-E_{\rm LSD}]/N$ in finite systems can be
extrapolated smoothly to infinite size~\cite{HRS}.) 
Thus the LDM equation for the closed-shell clusters, including the smaller 
ones, is better written as the difference to LSD~\cite{apf}: 
\begin{eqnarray}\label{LDM}
&\frac{E}{N}-\frac{E^{LSD}}{N}= 
(\epsilon^{unif}-\epsilon^{unif}_{PW92})  \label{eqdiffldmn} \\
&+ 4\pi r_s^2 (\sigma-\sigma^{LSD}) N^{-1/3} + 
2\pi r_s (\gamma-\gamma^{LSD}) N^{-2/3}.\nonumber
\end{eqnarray}
As the functionals PBE, PKZB and TPSS have the same 
parametrization (PW92)~\cite{PW92} in the limit of the uniform 
electron gas, the first term in the right-hand side is zero.

We used surface energies calculated by a planar 
surface code of Monnier and Perdew \cite{MP78} and reported in 
Table~\ref{surfxc}. Thus only the curvature-energy term in 
Eq.~({\ref{LDM}}) needs to be fitted.
To perform this fit we extended the calculation of jellium spheres up to
$N=$ 1100 (except in the case of $r_s$=5.62, where we only reached up to
$N=$748). We checked the ``Aufbau" principle, {\it i.e.}, the highest
occupied Kohn-Sham orbital should have an energy lower than the lowest
unnoccupied orbital, and we only took the clusters obeying that principle.
An example of such fitting is plotted in Fig.~\ref{fit400}.

\begin{table}
\caption{Surface exchange-correlation energies of jellium calculated with a
planar surface code. Energies are in erg/cm$^2$. Only deviations 
from LSD are relevant to fitting Eq.~({\ref{LDM}}).}
\begin{tabular}{ccccc}
\hline \hline
~~$r_s$~~       &       ~~~LSD~~~       &       ~~~PBE~~~       &       ~~~PKZB~~~      &       ~~~TPSS~~~      \\
\hline
2.07    &       2961    &       2881    &       3002    &       2985    \\
2.65    &       1204    &       1167    &       1221    &       1215    \\
3.24    &       575.1   &       555.9   &       583.1   &       581.8   \\
3.93    &       279.8   &       269.9   &       283.8   &       284.0   \\
5.62    &       69.89   &       67.27   &       71.15   &       71.90   \\
\hline
\hline
\end{tabular}
\label{surfxc}
\end{table}
\begin{table}
\caption{Curvature total energies of jellium
(in units of millihartree/bohr) of jellium.
Only deviations from LSD are relevant to fitting Eq.~({\ref{LDM}}).}  
\begin{tabular}{ccccc}
\hline \hline
~~~$r_s$~~~~    &       ~~~LSD~~~       &       ~~~PBE~~~       &       ~~~PKZB~~~ & ~~~TPSS~~~ \\
\hline
2.07    &       1.830   &       1.494   &       0.988   &       1.063   \\
2.65    &       1.044   &       0.885   &       0.548   &       0.609   \\
3.24    &       0.635   &       0.546   &       0.312   &       0.358   \\
3.93    &       0.369   &       0.318   &       0.156   &       0.189   \\
5.62    &       0.180   &       0.161   &       0.082   &       0.097   \\
\hline
\hline
\end{tabular}
\label{curvatures}
\end{table}

Using the resulting fit to curvature-energy differences and the
LSD curvature energies given by Ziesche {\it et al.} \cite{zpf},
we show in Table \ref{curvatures} our curvature energies for several
densities.
The energies are in fact very similar to a previous calculation \cite{apf}, 
which used only a smaller number of spheres and  did not restrict the 
surface energy term.
The curvature energies of TPSS are close to those of PKZB, as expected. 
The PKZB curvature energies are smaller than those of LSD and PBE, 
as predicted in Ref.~\onlinecite{apf}. 

\begin{figure}
\includegraphics[width=\columnwidth]{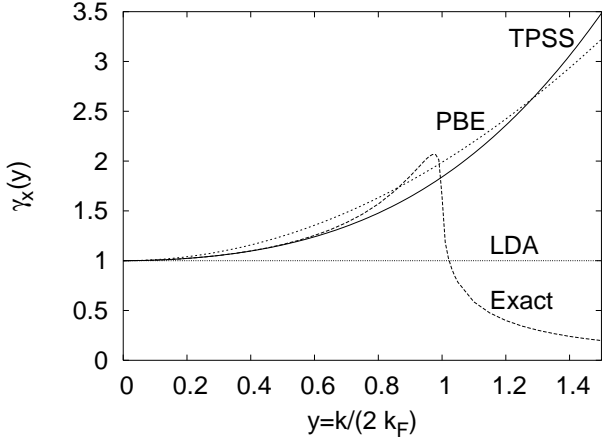}
\caption{The exchange-only response function $\gamma_\mathrm{x}$ 
for bulk jellium in the TPSS (full
line), PBE (short dashed) and LSD (dotted) functionals. The long dashed line
shows the exact-exchange only results from
Ref.~{\protect\onlinecite{antkl}}. See text
for discussion.}
\end{figure}
\begin{figure}
\includegraphics[width=\columnwidth]{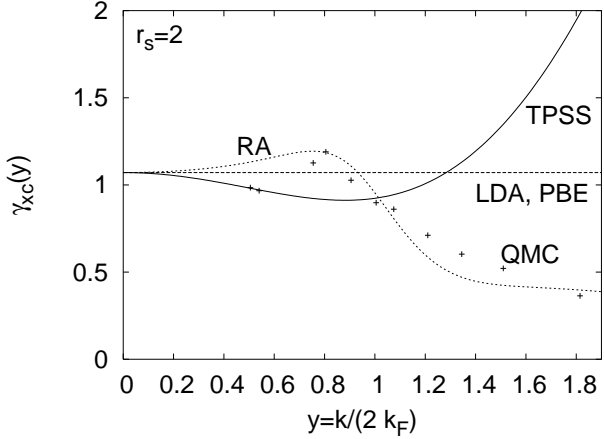}
\caption{The exchange-correlation response function $\gamma_\mathrm{xc}$ 
for bulk jellium as obtained
from the TPSS functional (full line), the LSD and PBE functionals (long
dashed), the Richardson-Ashcroft (RA) approximation\cite{ra} (short dashed) and the
Quantum Monte Carlo calculations of
Ref.~{\protect\onlinecite{moroni}} (crosses) for $r_s=2$.}
\end{figure}
\begin{figure}
\includegraphics[width=\columnwidth]{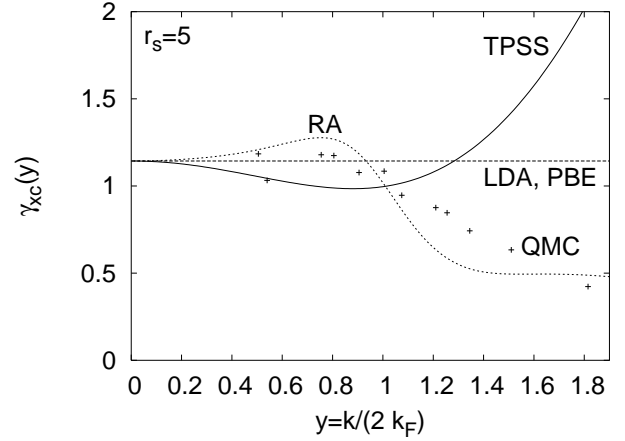}
\caption{Same as Fig. 8 but for a density with $r_s=5$.}
\end{figure}
\section{Linear density response and charge density waves}

There are several reasons why the linear response of the homogenous electron
gas to an external perturbation is of interest. First, it is an important step
towards a qualitative understanding of the electronic structure of the simple
metals, via pseudopotential perturbation theory~\cite{Harrison}. 
Second, comparing the response obtained from approximate density
functionals to the one obtained from Quantum Monte Carlo calculations can
serve as a test of different density
functionals. Third, the linear response relations can be used to study the
instability of the uniform phase of jellium against the formation of a 
charge-density wave~\cite{pd}.

In the present work, we calculate the linear response for LSD
\cite{PW92}, PBE GGA \cite{PBE}, and the PKZB \cite{PKZB} and
TPSS \cite{TPSS} meta-GGAs.
The quantity of interest is the response function
$\gamma_\mathrm{xc}(\mathbf{q})$ defined by the linear response relation
\begin{equation}
\label{defgam}
\delta
v_\mathrm{xc}(\kv)=- \frac{\pi}{k_\mathrm{F}^2}
\gamma_\mathrm{xc}\left(\frac{k}{2 \kf}\right) \delta n(\kv).
\end{equation}
This equation relates the Fourier component $\delta n(\kv)$ of a
density perturbation to the Fourier component $\delta
v_\mathrm{xc}(\kv) $ of the exchange-correlation potential that results from
the perturbed density. We calculated $\gamma_{\rm xc}$ 
separately for the exchange and
correlation parts of the different functionals by inserting the perturbed
density $n(\re)=n_0+n_\kv \cos(\kv \re)$, where $n_0=\kf^3/(3 \pi^2)$
and $n_\kv \ll n_0$, into the energy functional. The
resulting expression can be rewritten as a power series in $n_\kv/n_0$ and
$\gamma$ is obtained by multiplying the
second-order coefficient of this expansion by $-\frac{2 \kf^2}{\pi}$. 
(The factor 2 takes care of the 1/2 in the Taylor expansion.)
This procedure straightforwardly yields $\gamma_\mathrm{x}^\mathrm{LSD}=1$ and
$\gamma_\mathrm{x}^\mathrm{PBE}=1+\frac{9}{2}\mu y^2$, where $y=\frac{k}{2
  \kf}$ and $\mu=\beta \pi^{2/3}$, with $\beta=0.066725$ being 
the coefficient of the second-order gradient term in the gradient expansion
of the correlation energy in the high-density limit~\cite{MB68}.

For the meta-GGA
functionals the situation is more complicated, because they also depend on
the kinetic energy density $\tau$, which must be calculated from the
orbitals. We can, however, obtain the linear response without knowing the
orbitals in certain limiting cases by
using the appropriate gradient expansions of the kinetic energy density.
For a slowly-varying perturbation, i.e., $k\rightarrow 0$, the gradient
expansion reads \cite{brackjennchu}
\begin{equation}
\tau\approx \frac{3}{10}(3\pi^2)^{2/3}n^{5/3}+
\frac{1}{6}\nabla^2 n +
\frac{1}{360}\frac{\nabla^4 n}{(3 \pi^2)^{2/3}n^{2/3}}.
\end{equation}
We here retained only terms up to order $n_\kv$ since higher orders will not
contribute to the linear response. Inserting this expression into the
TPSS energy functional, Eq.(3) of Ref.\ \onlinecite{TPSS}, and again expanding into
a series in
$n_\kv/n_0$, we obtain after some algebra the slowly-varying limit of the TPSS
linear response function
\begin{equation}
\label{gamslow}
\gamma_\mathrm{x,slow}^\mathrm{TPSS}=1+\frac{5}{9}y^2+\frac{73}{225}y^4 -
\frac{146}{3375}y^6 + \mathcal{O}(y^8).
\end{equation}

In the rapidly-varying limit, i.e., $\kv \rightarrow \infty$
and $n_\kv \ll 1$, the kinetic energy density can be expanded as~\cite{JY71}
\begin{equation}
\label{taurapvar}
\tau \approx \frac{3}{10}(3\pi^2)^{2/3}n^{5/3}+
\frac{1}{8}\frac{|\nabla n|^2}{n}
+ c \nabla^2 n.
\end{equation}
To the best of our knowledge, no value for the coefficient $c$ has been given
in the past. We argue that $c=0$ for the following
reason: It is known that the von Weizs\"acker kinetic energy density is a
rigorous lower bound for the kinetic energy density,
\begin{equation}
\label{tauin}
\tau \ge \frac{1}{8}\frac{|\nabla n|^2}{n}.
\end{equation}
But for $c\ne 0$ and any non-vanishing amplitude of the perturbation, one can
choose a wavevector to make the $\nabla^2$-term in
Eq.(\ref{taurapvar}) arbitrarily large and negative for certain points in
space. Thus, the only way to avoid the violation of the inequality (\ref{tauin}) 
is to simply require $c=0$.
Using Eq.(\ref{taurapvar}) with $c=0$ yields the rapidly-varying limit of
TPSS linear response,
\begin{equation}
\label{gamrap}
\gamma_\mathrm{x,rapid}^\mathrm{TPSS}=1+\frac{5}{9}y^2.
\end{equation}

Finally interpolating between the two limiting cases of Eq.(\ref{gamslow}) and
Eq.(\ref{gamrap}) with the Pad\'{e} approximant
\begin{equation}
\gamma_\mathrm{x}^\mathrm{P,TPSS}=1+\frac{5}{9}y^2+
\frac{\frac{73}{225}y^4}{(1+\frac{2}{45}y^2)^3},
\end{equation}
we obtain an expression that can be
expected to be very close to the exact TPSS response
$\gamma_\mathrm{x}^\mathrm{TPSS}$ for all k.
Going through the same procedure for the PKZB meta-GGA we confirmed that PKZB
and TPSS have the same linear response.

In Fig. 7 we plotted the exchange-only response functions for LSD, PBE GGA, 
and TPSS meta-GGA,
together with the exact exchange-only response from Ref. \onlinecite{antkl}. The
TPSS exchange-only response is extremely close to the exact exchange-only
response up to $y\approx 0.6$, and both PBE and TPSS provide reasonable
approximations up to $y\approx 1$. Only for wavevectors with a magnitude of
more than twice the Fermi wavenumber do differences become pronounced since
the semi-local functionals do not recover the abrupt drop of the exact
exchange response beyond $y\approx 1$.

To calculate the correlation contribution to the LSD linear response, we
proceed slightly differently and directly evaluate
$\gamma_\mathrm{c}^\mathrm{LSD}=-(\kf^2/\pi)
\delta^2 E_\mathrm{c}^\mathrm{LSD}/\delta n(\re)\delta n(\rp)
$
with
the Perdew-Wang expression \cite{PW92}. Obviously,
$
\delta^2 E_\mathrm{c}^\mathrm{LSD}/\delta n(\re)\delta n(\rp)=
\left[2\partial \epsilon_\mathrm{c}^\mathrm{uni}/\partial n+
n\partial^2 \epsilon_\mathrm{c}^\mathrm{uni}/\partial n^2
 \right]\delta(\re-\rp),
$
and some algebra yields
\begin{equation}
\frac{\partial \epsilon_\mathrm{c}^\mathrm{uni}}{\partial n}=
-\frac{4\pi r_s^4}{9}A\left[\frac{c_1 q_2}{\sqrt{r_s}c_3 q_1}-
2\alpha_1 \ln\left( 1+\frac{1}{c_2}\right)   \right]
\end{equation}
and
\begin{eqnarray}
\frac{\partial^2 \epsilon_\mathrm{c}^\mathrm{uni}}{\partial n^2}&=&
\frac{16}{81}\pi^2r_s^7
[c_1 q_2^2 + 4 \alpha_1 q_1 q_2 c_3 \sqrt{r_s} - 2 c_1 q_2^2 c_3
\nonumber \\
    &+& c_1 c_3 (-\beta_1+3\beta_3 r_s +8 \beta_4 r_s^{3/2} )/\sqrt{r_s}
] \nonumber \\&&
/\left[8 A q_1^4 (1+1/c_2)^2\right] \nonumber \\
&+&
4\left[\frac{Ac_1 q_2}{\sqrt{r_s}q_1 c_3}-2 A \alpha_1 \ln(1+1/c_2)\right].
\end{eqnarray}
Here 
$q_1=\beta_1 \sqrt{r_s} + \beta_2 r_s + \beta_3 r_s^{3/2} + \beta_4 r_s^2$,
$q_2=\beta_1 + 2 \beta_2  \sqrt{r_s} + 3 \beta_3 r_s + 4 \beta_4 r_s^{3/2}$,
$c_1=1+\alpha_1 r_s$,
$c_2=2 a q_1$,
and 
$c_3=1+c_2$,
with $A$, $\alpha_1$, and $\beta_1$--$\beta_4$ being the parameters from Ref.\
\onlinecite{PW92}.

The contribution of the correlation part to the
response for the PBE GGA and the two MGGAs is obtained analogously to the
calculations for the exchange functionals. In all three cases we obtain the
same result,
\begin{equation}
\gamma_\mathrm{c}^\mathrm{PBE}=
\gamma_\mathrm{c}^\mathrm{PKZB}=
\gamma_\mathrm{c}^\mathrm{TPSS}=\gamma_c^{\rm LSD}-\frac{3}{2}\beta \pi^2 y^2,
\end{equation}
where $\beta$ is the same parameter as in the PBE exchange response function
(see above).

In Figs. 8 and 9 we compare the response as obtained from the density
functionals to the Quantum Monte Carlo results of Ref. \onlinecite{moroni} and the
results of Ref. \onlinecite{ra} for two different densities.
LSD and PBE provide a satisfying average $\gamma$ for $y<1$. The TPSS response
shows more structure than LSD and PBE and is also in satisfactory agreement
with the QMC results. Only for $y>1.1$ does the TPSS response move outside of
the QMC error bars.

Finally, following Ref. \onlinecite{pd} we calculated the instability of
jellium against
the formation of a charge density wave. The Fourier components of the
self-consistent potential are
related to the ones of the external potential by
\begin{equation}
v(\kv)=\frac{1}{1-\frac{4 \pi}{k^2}[1-y^2 \gamma_\mathrm{xc}(y)]\chi(k)}
v_\mathrm{ext}(\kv),
\end{equation}
where $\chi$ is the Lindhard response function. 
For vanishing $v_\mathrm{ext}(\kv)$,
the left-hand side of this equation can only be nonzero if the denominator on
the right-hand side vanishes. Thus, for each value of $y$ we numerically search
for the density (i.e., $\kf$) that sets the denominator to zero. The largest value
of $\kf$ (for all $y$) found in that way marks the onset of jellium
instability. In this way we
confirmed that, for exchange and correlation combined, 
TPSS does not alter much the prediction of LSD for the 
onset of jellium instability, which occurs for 
$\kf < 0.06$ a.u. ($r_s \gtrsim 30$).

\section{Conclusions}
In conclusion, we have calculated and compared the jellium surface 
exchange-correlation
energies of the PBE GGA and the TPSS meta-GGA and of these two functionals
when uniformly scaled to the high- and low-density limits for the normal
bulk valence densities in magnetic fields. In all the cases, the fairly 
good agreement of the PBE GGA with the TPSS meta-GGA shows that the TPSS
meta-GGA indeed represents the self-correlation correction of the PBE GGA. 
We have further found that the ``internal'' magnetic field of 
Eq.(\ref{Kautz}) and the external uniform field of Eq.(\ref{ew}) 
are typically opposite to each 
other in their effects on the work function and surface 
exchange-correlation energy of jellium.

We have also calculated the energies of jellium spheres with LSD, 
PBE GGA, and TPSS and its predecessor PKZB meta-GGAs.
Typically, while PBE energies are too low for spheres with 
more than about two electrons, LSD and TPSS are accurate there, 
up to 106 electrons. Curvature energies are 
reduced substantially as we pass from LSD to PBE to TPSS. Finally,
we have shown that the linear response of bulk jellium (to perturbations 
with wavevectors less than twice the Fermi wavevector) is 
reasonably described by all the functionals considered here. 

As we climb the ladder of nonempirical density functional 
approximations from LSD to GGA to meta-GGA, there is a steady and 
dramatic improvement in atomization energies~\cite{SSTP1,FP06}. Surface
energies worsen~\cite{PP,SSTP2,CPT,PCP,W07,ybs06,CPDGP,PKZB} 
from LSD to PBE GGA (and other 
popular GGAs), due to an imperfect error cancellation between exchange and
correlation, but this can be corrected in any of three ways: 
(1) by transferring~\cite{MM} the needed correction from jellium to
real systems, (2) by using GGAs designed specifically for solids 
(and not for free atoms)~\cite{ann05,WC06,pbesol}, or (3) by 
climbing up further to the TPSS meta-GGA.  The third way adds 
little in computational cost~\cite{SSTP1,FP06}, even at 
full selfconsistency, and seems worthy 
of further testing and possible refinement.

The jellium model itself remains useful as a testing ground 
for density functionals.  Although some of its properties become 
unphysical as one moves away from the bulk density at which
jellium is stable ($r_s \approx 4$, roughly the valence density of sodium), 
this problem can also be fixed inexpensively via the stabilized 
jellium model~\cite{PTS, zpf}.

\section*{ACKNOWLEDGMENTS}
This work was supported by the NSF under Grants DMR-0135678 (J.P.P. and J.T.) 
and DMR-0501588 (J.P.P.), by DOE under Contract No. DE-AC52-06NA25396 and
Grant No. LDRD-PRD X9KU at LANL (J.T.), and 
by the Deutsche Forschungsgemeinschaft (S.K.).


\begin{thebibliography}{100}
\bibitem{LK70}
N.D. Lang and W. Kohn, Phys. Rev. B {\bf 1}, 4555 (1970).
\bibitem{KS}
W. Kohn and L. J. Sham, Phys. Rev. {\bf 140}, A1133 (1965).
\bibitem{PK}
J.P. Perdew and S. Kurth, in {\it A Primer in Density Functional Theory},
edited by C. Fiolhais, F. Nogueira, and M. Marques,
Lecture Notes in Physics {\bf 620} (Springer, Berlin, 2003).
\bibitem{WK}
W. Kohn, Rev. Mod. Phys. {\bf 71}, 1253 (1999).
\bibitem{Mahan}
G.D. Mahan, Phys. Rev. B {\bf 12}, 5585 (1975).
\bibitem{PM76}
J.P. Perdew and R. Monnier, Phys. Rev. Lett. {\bf 37}, 1286 (1976).
\bibitem{P77}
J.P. Perdew, Phys. Rev. B {\bf 16}, 1525 (1977).
\bibitem{MP78}
R. Monnier and J.P. Perdew, Phys. Rev. B {\bf 17}, 2595 (1978). 
\bibitem{KS76}
R.L. Kautz and B.B. Schwartz, Phys. Rev. B {\bf 14}, 2017 (1976).
\bibitem{PP}
J.M. Pitarke and J.P. Perdew, Phys. Rev. B {\bf 67}, 045101 (2003).
\bibitem{SSTP2}
V.N. Staroverov, G.E. Scuseria, J. Tao, and J.P. Perdew,
Phys. Rev. B {\bf 69}, 075102 (2004).
\bibitem{CPT}
L.A. Constantin, J.P. Perdew, and J. Tao, Phys. Rev. B {\bf 73}, 205104 (2006).
\bibitem{PCP} 
J.M. Pitarke, L.A. Constantin, and J.P. Perdew, Phys. Rev. B {\bf 74}, 
045121 (2006).
\bibitem{W07}
B. Wood, N.D.M. Hine, W.M.C. Foulkes, and P. Garc\'ia-Gonz\'alez,
Phys. Rev. B {\bf 76}, 035403 (2007).
\bibitem{ybs06}
D.K. Yu, H.P. Bonzel, and M. Scheffler, Phys. Rev. B {\bf 74}, 115408 (2006).
\bibitem{CPDGP}
L.A. Constantin, J.M. Pitarke, J.F. Dobson, A. Garcia-Lekue, and 
J.P. Perdew, Phys. Rev. Lett. {\bf 100}, 036401 (2008).
\bibitem{PW86}
J.P. Perdew and Y. Wang, Phys. Rev. B {\bf 33}, 8800 (1986);
J.P. Perdew, Phys. Rev. B {\bf 33}, 8822 (1986); {\bf 34}, 7406 (1986)(E).
\bibitem{PBE}
J.P. Perdew, K. Burke, and M. Ernzerhof, Phys. Rev. Lett. {\bf 77},
3865 (1996).
\bibitem{SSTP1}
V.N. Staroverov, G.E. Scuseria, J. Tao, and J.P. Perdew,
J. Chem. Phys. {\bf 119}, 12129 (2003); ibid. {\bf 121}, 11507 (2004) (E).
\bibitem{KPB}
S. Kurth, J.P. Perdew, and P. Blaha, Int. J. Quantum Chem. {\bf 75},
889 (1999).
\bibitem{Perdew85}
J.P. Perdew, Phys. Rev. Lett. {\bf 55}, 1665 (1985).
\bibitem{PKZB}
J.P. Perdew, S. Kurth, A. Zupan, and P. Blaha, Phys. Rev. Lett. {\bf 82},
2544 (1999).
\bibitem{TPSS}
J. Tao, J.P. Perdew, V.N. Staroverov, and G.E. Scuseria,
Phys. Rev. Lett. {\bf 91}, 146401 (2003).
\bibitem{SB96}
P.S. Svendsen and U. von Barth, Phys. Rev. B {\bf 54}, 17402 (1996).
\bibitem{AES}
C. Adamo, M. Ernzerhof, and G.E. Scuseria, J. Chem. Phys. {\bf 112},
2643 (2000).
\bibitem{RS}
A.D. Rabuck and G.E. Scuseria, Theor. Chem. Acc. {\bf 104}, 439 (2000).
\bibitem{PTSS}
J.P. Perdew, J. Tao, V.N. Staroverov, and G.E. Scuseria,
J. Chem. Phys. {\bf 120}, 6898 (2004).
\bibitem{tprscs}
J. Tao, J.P. Perdew, A. Ruzsinszky, G.E. Scuseria, G.I. Csonka, and V.N. Staroverov,
Phil. Mag. {\bf 87}, 1071 (2007).
\bibitem{SSPT}
V.N. Staroverov, G.E. Scuseria, J.P. Perdew, J. Tao, and E.R. Davidson,
Phys. Rev. A {\bf 70}, 012502 (2004).
\bibitem{SP}
M. Seidl and J.P. Perdew, Phys. Rev. B {\bf 50}, 5744 (1994).
\bibitem{ZTPS}
P. Ziesche, J. Tao, M. Seidl, and J.P. Perdew,
Int. J. Quantum Chem. {\bf 77}, 819 (2000).
\bibitem{SPL}
M. Seidl, J.P. Perdew, and M. Levy, Phys. Rev. A {\bf 59}, 51 (1999).
\bibitem{SPK}
M. Seidl, J.P. Perdew, and S. Kurth, Phys. Rev. A {\bf 62}, 012502 (2000).
\bibitem{March1}
L. Miglio, M.P. Tosi, and N.H. March, Sur. Sci. {\bf 111}, 119 (1981).
\bibitem{Perdew2001}
J.P. Perdew and K. Schmidt,
\newblock in {\em Density Functional Theory and Its Application to Materials},
  edited by V.~{Van Doren}, C.~{Van Alsenoy}, and P.~Geerlings (AIP, Melville,
  New York, 2001).
\bibitem{prtssc}
J.P. Perdew, A. Ruzsinszky, J. Tao, V.N. Staroverov, G.E. Scuseria,
G.I. Csonka, J. Chem. Phys. {\bf 123}, 062201 (2005).
\bibitem{ann05}
R. Armiento and A.E. Mattsson, Phys. Rev. B {\bf 72}, 085108 (2005).
\bibitem{WC06}
Z. Wu and R.E. Cohen, Phys. Rev. B {\bf 73}, 235116 (2006).
\bibitem{pbesol}
J.P. Perdew, A. Ruzsinszky, G.I. Csonka, O.A. Vydrov, G.E. Scuseria, 
L.A. Constantin, X. Zhou, and K. Burke, Phys. Rev. Lett. {\bf 100},
136406 (2008).
\bibitem{OP}
G.L. Oliver and J.P. Perdew, Phys. Rev. A {\bf 20}, 397 (1979).
\bibitem{LP85}
M. Levy and J.P. Perdew, Phys. Rev. A {\bf 32}, 2010 (1985).
\bibitem{PTK}
J.P. Perdew, J. Tao, and S. K\"ummel, in {\it Electron
Correlation Methodology}, edited by A.K. Wilson and K.A. Peterson
(ACS Symposium Series 958, distributed by Oxford University Press, 2007).
\bibitem{PW92}
J.P. Perdew and Y. Wang, Phys. Rev. B {\bf 45}, 13244 (1992).
\bibitem{PBY96}
J.P. Perdew, K. Burke, and Y. Wang, Phys. Rev. B {\bf 54}, 16533 (1996).
\bibitem{WP91}
Y. Wang and J.P. Perdew, Phys. Rev. B {\bf 43}, 8911 (1991).
\bibitem{TAO01}
J. Tao, J. Chem. Phys. {\bf 115}, 3519 (2001).
\bibitem{apf} L.M. Almeida, J.P. Perdew, and C. Fiolhais,
Phys. Rev. B, 66,  075115 (2002).
\bibitem{MB68}
S.-K. Ma and K.A. Brueckner, Phys. Rev. {\bf 165}, 18 (1968).
\bibitem{GR76}
D.J.W. Geldart and M. Rasolt, Phys. Rev. B {\bf 13}, 1477 (1976).
\bibitem{LP80}
D.C. Langreth and J.P. Perdew, Phys. Rev. B {\bf 21}, 5469 (1980).
\bibitem{SGEL} 
M. Swart, A.R. Groenhof, A.W. Ehlers, and K. Lammertsma, 
J. Phys. Chem. {\bf 108}, 5479 (2004).
\bibitem{ZBFCC}
S. Zein, S.H. Borshch, P. Fleurat-Lessard, M.E. Casida, and H. Chermette, 
J. Chem. Phys. {\bf 126}, 014105 (2007).
\bibitem{Swart}
M. Swart, private communication.
\bibitem{Bardeen}
J. Bardeen, Phys. Rev. {\bf 49}, 653 (1936).
\bibitem{March2}
I.D. Moore and N.H. March, Ann. Phys. {\bf 97}, 136 (1976).
\bibitem{LP00}
L. Pollack and J.P. Perdew, J. Phys-Condens. Mat. {\bf 12}, 1239 (2000).
\bibitem{YPK}
Z. Yan, J.P. Perdew, and S. Kurth, Phys. Rev. B {\bf 61}, 16430 (2000).
\bibitem{VS1}
V. Sahni, J. Gruenebaum, and J.P. Perdew, Phys. Rev. B {\bf 26},
4371 (1982).
\bibitem{sb} F. Sottile and P. Ballone, Phys. Rev. B {\bf 64}, {045105} (2001). 
\bibitem{ob94}  G. Ortiz and P. Ballone, Phys. Rev. B {\bf 50}, {1391} (1994).
\bibitem{ca80}  D.M. Ceperley and B.J. Alder,
Phys. Rev. Lett. {\bf 45}, A {566} (1980).
\bibitem{pwe} J.P. Perdew, Y. Wang, and E. Engel,
Phys. Rev. Lett. {\bf 66}, {508} (1991).
\bibitem{HRS}
Q.M. Hu, K. Reuter, and M. Scheffler, Phys. Rev. Lett, {\bf 99}, 
169903(E) (2007).
\bibitem{zpf} P. Ziesche, J.P. Perdew, and C. Fiolhais,
Phys. Rev. B, 49,  7916 (1994).
\bibitem{Harrison}
W.A. Harrison, {\it Pseudopotentials in the Theory of Metals}
(Benjamin, New York, 1966).
\bibitem{pd}
J.P. Perdew and T. Datta, Phys. Stat. Sol (b) {\bf 102}, 283 (1980).
\bibitem{brackjennchu}
M. Brack, B.K. Jennings, and Y.H. Chu, Phys. Lett. {\bf 65B}, 1 (1976).
\bibitem{JY71}
W. Jones and W.H. Young, J. Phys. C: Solid State Phys. {\bf 4}, 1322 (1971).
\bibitem{antkl}
P.R. Antoniewicz and L. Kleinman, Phys. Rev. B {\bf 31}, 6779 (1985).
\bibitem{moroni}
S. Moroni, D.M. Ceperley, and G. Senatore, Phys. Rev. Lett. {\bf 75}, 689 (1995).
\bibitem{ra}
C.F. Richardson and N.W. Ashcroft, Phys. Rev. B {\bf 50}, 8170 (1994).
\bibitem{FP06}
F. Furche and J.P. Perdew, J. Chem. Phys. {\bf 124}, 044103 (2006).
\bibitem{MM}
T.R. Mattsson and A.E. Mattsson, Phys. Rev. B {\bf 66}, 214110 (2002).
\bibitem{PTS}
J.P. Perdew, H.Q. Tran, and E.D. Smith, Phys. Rev. B {\bf 42}, 11627 (1990).
\end{thebibliography}
\end{document}